\documentclass[prb,amsmath,superscriptaddress,citeautoscript,twocolumn,
showpacs,floatfix]{revtex4}
\usepackage{bm}
\usepackage{amssymb}
\usepackage{graphicx}
\usepackage{amsmath, amsthm, amssymb, graphicx}

\newcommand{\ep}{\varepsilon}
\newcommand{\be}{\begin{equation}}
\newcommand{\ed}{\end{equation}}
\newcommand{\ber}{\begin{eqnarray}}
\newcommand{\edr}{\end{eqnarray}}
\newcommand{\bern}{\begin{eqnarray*}}
\newcommand{\edrn}{\end{eqnarray*}}
\newcommand{\bfl}{\begin{flalign}}
\newcommand{\efl}{\end{flalign}}

\begin{document}

\title{Pure Phase Decoherence in a Ring Geometry }

\author{Z. Zhu}
\affiliation{Department of Physics and Astronomy, University
of British Columbia, 6224 Agricultural Rd., Vancouver, B.C., Canada
V6T 1Z1}

\author{A. Aharony}

\altaffiliation{Also at Tel Aviv University,
Tel Aviv 69978, Israel}

\affiliation{Department of Physics and the Ilse Katz Center for
Meso- and Nano-Scale Science and Technology, Ben Gurion
University, Beer Sheva 84105, Israel}

\affiliation{Pacific
Institute of Theoretical Physics, University of British Columbia,
6224 Agricultural Rd., Vancouver, B.C., Canada V6T 1Z1}

\author{O. Entin-Wohlman}

\altaffiliation{Also at Tel Aviv University,
Tel Aviv 69978, Israel}

\affiliation{Department of Physics and the Ilse Katz Center for
Meso- and Nano-Scale Science and Technology, Ben Gurion
University, Beer Sheva 84105, Israel}

\affiliation{Pacific
Institute of Theoretical Physics, University of British Columbia,
6224 Agricultural Rd., Vancouver, B.C., Canada V6T 1Z1}

\author{ P.C.E. Stamp}
\affiliation{Department of Physics and Astronomy, University
of British Columbia, 6224 Agricultural Rd., Vancouver, B.C., Canada
V6T 1Z1}
 \affiliation{Pacific
Institute of Theoretical Physics, University of British Columbia,
6224 Agricultural Rd., Vancouver, B.C., Canada V6T 1Z1}
\date{{\small \today}}

\begin{abstract}
We study the dynamics of pure phase decoherence for a particle
hopping around an $N$-site ring, coupled both to a spin bath and to
an Aharonov-Bohm flux which threads the ring. Analytic results are
found for the dynamics of the influence functional and of the
reduced density matrix of the particle, both for initial single
wave-packet states, and for states split initially into 2 separate
wave-packets moving at different velocities. We also give results
for the dynamics of the current as a function of time.
\end{abstract}


\maketitle


\section{ Introduction}
 \label{sec:intro}


The  dynamics of phase decoherence is central to our understanding
of those physical systems whose properties depend on interference.
This is particularly evident when particles are forced to propagate
around closed paths; phase coherence then makes all physical
properties depend on the topology of these paths \cite{thouless}.
For this reason the quantum dynamics of particles on rings has been
extremely important in our understanding of quantum phase coherence.
Examples at the microscopic level include the energetics and
response to magnetic fields of molecules \cite{orbital}, as well as
charge transfer dynamics in a vast array of solid-state and
biochemical systems. There is evidence now for coherent transport
around ring structures even in some large biomolecules \cite{LH-2}.
At the nanoscopic and mesoscopic scale many ring-like structures,
both conducting and superconducting \cite{AhB-scond}, show coherent
transport around the rings, along with interesting Aharonov-Bohm
style interference phenomena. We also note the importance of closed
loop structures in quantum information processing \cite{QIP}.

The interference around loops in all of these systems is very
sensitive to phase decoherence. Questions about the mechanisms and
dynamics of this decoherence are subtle, and have led to major
controversies, notably in the discussion of mesoscopic conductors
\cite{tau-phi}. A quantitative understanding of decoherence
processes in metallic systems and in superconducting "qubits" has
yet to be attained (in both cases local defect modes clearly make
the major contribution to phase decoherence at low temperature $T$
\cite{TLS-TauPhi,TLS-SQUID}). These controversies are examples of a
wider problem: typically in solid-state systems, low $T$ decoherence
rates are far higher in experiments than theoretical estimates based
on the dissipation rates in these systems.

These problems are complex because both decoherence and dissipation
rates depend strongly on which environmental modes are causing the
decoherence \cite{SHPMP,PS00}. Delocalized modes (electrons,
phonons, photons, spin waves, etc.) can typically be modeled as
"oscillator bath" modes \cite{feynman63,cal83,weiss99}. In such
models, decoherence goes hand-in-hand with dissipation
\cite{AJL84,cal84}, in accordance with the fluctuation-dissipation
theorem. However localized modes (defects, dislocations, dangling
bonds, nuclear and paramagnetic impurity spins, etc.), which can be
mapped to a "spin bath" representation of the environment
\cite{SHPMP,PS00}, behave quite differently; indeed they often give
decoherence with almost no dissipation. This is because although
their low characteristic energy scale means they can cause little
dissipation, nevertheless their phase dynamics can be strongly
affected when they couple to some collective coordinate - this then
causes strong decoherence in the dynamics of this coordinate
\cite{PS00,gaita08}. The fluctuation-dissipation theorem is then not
obeyed \cite{SHPMP}, and often these localized modes are rather far
from equilibrium.

\begin{figure}[h]
\includegraphics[width=8cm]{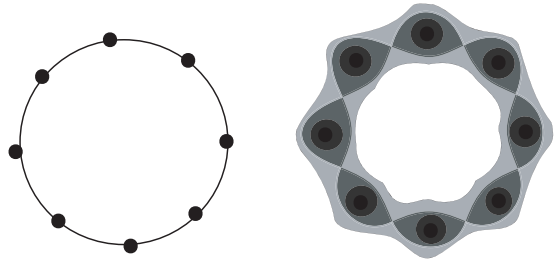}
\caption{\label{fig:RING}At left, An 8-site ring with
nearest-neighbour hopping between sites. At right a potential
$U({\bf R})$ with 8 potential wells (shown here symmetric under
rotations by $\pi/4$), depicted as a contour map (with lower
potential shown darker). When truncated to the 8 lowest eigenstates,
this is equivalent to the 8-site model.}
\end{figure}

To understand how such non-dissipative decoherence processes work,
it is then useful to look at models in which the environment causes
pure phase decoherence, with no dissipation. As noted above, such
models become particularly interesting when the decoherence is
acting on systems propagating in 'closed loops'. Models of rings
coupled to oscillator baths have already been studied \cite{guinea}.
However such models, in which decoherence is inextricably linked to
dissipation, do not capture the largely non-dissipative decoherence
processes that dominate many solids at low $T$. On the the other
hand pure phase decoherence has been studied in many
papers\cite{unruh95,ekert96,milburn05}, but not, as far as we know,
the rather unique phenomena occurring on a ring.

In this paper we study a model which embodies in a simple way both
the 'closed path' propagation which is generic to quantum
interference processes, and which involves pure phase decoherence
coming from a spin bath. The model describes a particle propagating
around a ring of $N$ discrete sites, while coupled to a spin bath;
we assume hopping between nearest neighbors. The model becomes
particularly interesting if we also have a flux $\Phi$ threading the
ring (see Fig. \ref{fig:RING}). The spin bath variables are assumed
to be Two-Level Systems (TLS); these are ubiquitous in solid-state
systems, and are the main cause of decoherence at low $T$ in these
systems.

One can also study the problem of a continuous ring, but the
discrete model is simpler, and is easily related to diverse problems
like quantum walks with phase decoherence \cite{PS06,QW-diss}, or
the dynamics of electrons in rings of quantum dots \cite{QDot-ring}.
The Hamiltonian we will study has the general form
 \begin{equation}
  \label{eq:H_phi'}
H_{\phi}\;=\;\sum_{<ij>}[\Delta_o c^{\dag}_i c_j e^{i(A_{ij}^o +
\sum_k\mbox{\boldmath $\alpha$}_{k}^{ij}\cdot \mbox{\boldmath
$\sigma$}_k)}+H.c.]
 \end{equation}
The operator $c_j^{\dag}$ creates a particle at site $j$; we assume
a single particle only. The phase factors $\{ A_{ij}^{o} \}$ result
from the flux $\Phi$ threading the ring. In writing
(\ref{eq:H_phi'}), we have assumed a symmetric ring, with $N$ sites,
and assumed that the hopping matrix elements $t_{ij}$ between sites
$i$ and $j$ have simplified to a nearest-neighbour amplitude
$\Delta_o$ (here $\sum_{<ij>}$ denotes a sum over nearest
neighbours).  This also means we can ignore any diagonal site
energies, since symmetry under rotations by angles $2\pi/N$ means
these energies are all the same. The spin bath variables $\{
\mbox{\boldmath $\sigma$}_k \}$ are Pauli spin-$1/2$ operators for
the TLS, with $k=1,2,....N_s$. We emphasize immediately that these
bath spins are, in real situations, often not spins, but instead the
2 lowest levels of localized modes in a solid (for example, as noted
above, they could be defects or dangling bonds).

The paper is organized as follows. In section II we discuss the
derivation of model Hamiltonians like (\ref{eq:H_phi'}) from more
microscopic models, and the approximations which allow us to drop
other terms that can also appear in the coupling of a ring particle
to a spin bath. In section III we discuss the dynamics of the
particle in the absence of the bath - this establishes a number of
useful mathematical results. In section IV we show how the dynamics
of the reduced density matrix for the particle is derived in the
presence of the bath, and give some results for this dynamics. In
section V we analyze the dynamics of a pair of interfering
wave-packets moving around the ring, showing how pure phase
decoherence destroys the interference between them. Finally, in
section VI, we summarize our conclusions - since some of the
calculations are quite extensive, readers may want to look first at
this section for a guide to the main results. The more technical
details of the derivations in sections III and IV are given in an
Appendix.


\section{ Derivation of Model}
 \label{sec:derivation}


Consider first an $N$-site ring system without a bath. In site
representation, this typically has a "bare ring" model Hamiltonian
 \begin{equation}
 \label{eq:H_o}
H_o \;=\; \sum_{<ij>} \left[ t_{ij} c_i^{\dagger} c_j \: e^{i
A_{ij}^{o}}  + H.c.\right] + \sum_j\ep_j c_j^{\dagger} c_j
 \end{equation}
This "1-band" Hamiltonian is the result of truncating, to low
energies, a high-energy Hamiltonian of form:
 \begin{equation}
H_V\;=\;\frac{1}{2 M}({\bf P} - {\bf A}({\bf R}))^2 +
U(\mbox{\boldmath $R$})
 \end{equation}
where a particle of mass $M$ moves in a potential $U(\mbox{\boldmath
$R$})$ characterized by $N$ potential wells in a ring array (see
again Fig. \ref{fig:RING}). Then $\ep_j$ is the energy of the lowest
state in the $j$-th well, and $t_{ij}$ is the tunneling amplitude
between the $i$-th and $j$-th wells (which we take here to be
nearest neighbours). In path integral language, this tunneling is
over a semiclassical "instanton" trajectory ${\bf R}_{ins} (\tau)$,
occurring over a timescale $\tau_B \sim 1/\Omega_0$ (the "bounce
time" \cite{coleman}). Here $\Omega_0$ (the "bounce frequency") is
roughly the small oscillation frequency of the particle in the
potential wells. In a semiclassical calculation, the phase
$A_{ij}^o$ is that incurred along the semiclassical trajectory by
the particle, moving in the gauge field ${\bf A}({\bf R})$. For a
symmetric ring the site energy $\ep_j\rightarrow \ep_0,\forall j$,
and we henceforth ignore it.

Consider now what happens when we couple the particle to a spin
bath. The spin bath itself, independent of the ring particle, has
the Hamiltonian
 \begin{equation}
 \label{eq:H_SB}
H_{SB} \;=\; \sum_{k}{\bf h}_k \cdot \mbox{\boldmath $\sigma$}_k
\;+\; \sum_{k,k'} V_{kk'}^{\alpha \beta} \sigma_{k}^{\alpha}
\sigma_{k'}^{\beta}
 \end{equation}
in which each TLS has some local field ${\bf h}_k$ acting on it, and
the interactions $V_{kk'}^{\alpha \beta}$ are typically rather small
because the TLS represent localized modes in the environment. The
most general coupling between the ring particle and the bath has the
form
\begin{eqnarray}
 H_{int} \;=\; &\sum_{k}^{N_s}&
[\; \sum_j \mbox{\boldmath $F$}_j^k(\mbox{\boldmath $\sigma$}_k)
\hat{c}_{j}^{\dagger}\hat{c}_j  \nonumber\\
 \;\;\;\;\;\;\;\;&+& \sum_{<ij>} (
\mbox{\boldmath $G$}_{ij}^k(\mbox{\boldmath $\sigma$}_k)
\hat{c}_{i}^{\dagger}\hat{c}_j +H.c.) ]
 \label{spin-z}
\end{eqnarray}
in which both the diagonal coupling $\mbox{\boldmath $F$}_j^k$ and
the non-diagonal coupling $\mbox{\boldmath $G$}_{ij}^k$ are vectors
in the Hilbert space of the $k$-th bath spin. We shall see below,
when considering the origin of these terms from microscopic models,
that very often we can write the total Hamiltonian as
 \begin{equation}
 \label{eq:H} H
\;=\; H_{band} + H_{SB}
 \end{equation}
where $H_{band} = H_o + H_{int}$ takes the form
\begin{eqnarray}
H_{band} = &\sum_{ij} [t_{ij} c_i^{\dagger} c_j  e^{i A_{ij}^{o} +
i\sum_k (\mbox{\boldmath $\phi$}_k^{ij}+ \mbox{\boldmath
$\alpha$}_{k}^{ij}\cdot \mbox{\boldmath $\sigma$}_k)} + H.c.]
\nonumber\\
 &+ \;\sum_j (\ep_j+\sum_k \mbox{\boldmath $\gamma$}_{k}^{j} \cdot
\mbox{\boldmath $\sigma$}_k) c_j^{\dagger} c_j
 \label{H-band1}
\end{eqnarray}
in which the diagonal couplings to the spin bath assume a "Zeeman"
form, of strength $|\mbox{\boldmath $\gamma$}_{k}^{j}|$, linear in
the $\{ \mbox{\boldmath $\sigma$}_k \}$, and the non-diagonal
couplings appear in the form of extra phase factors in the hopping
amplitude between sites.

Before we consider the microscopic origins of this model, let us
note how it simplifies when we assume the symmetry under rotations
by $2\pi/N$ noted above (so that the site energy $\ep_j$ is dropped,
and $t_{ij} \rightarrow \Delta_o$, with nearest-neighbour hopping
only). It is then natural to write $A_{ij}^o$ as
\begin{align}
A_{ij}^o = \frac{e}{2}{\bf H}\cdot{\bf R}_{i}^{}\times{\bf R}_{j}^{}
\;=\; \Phi/N
\end{align}
for $j=i+1$ (we now use MKS units, and put $\hbar =1$). Here, ${\bf H}$ is the
magnetic field, and ${\bf R}_{i}$ is the radius-vector to the $i$th
site; in cylindrical coordinates
\begin{align}
{\bf R}_j &= (R_o, \Theta_j)  \nonumber \\
\Theta_j &= 2 \pi j/N
 \label{theta-j}
\end{align}
for a ring of radius $R_o$. Fourier transforming from the site basis
to a momentum basis for the couplings, we define quasi-momenta $k_n
= 2\pi n/N$, with $n = 0,1,2,...,N-1$, for the particle on the ring,
and define operators
\begin{align}
c^{\dagger}_{j}&=\sqrt{\frac{1}{N}}\sum_{k^{}_{n}}e^{ik^{}_{n} j}
c^{\dagger}_{k^{}_{n}}\ ,\nonumber\\
c^{\dagger}_{k^{}_{n}}&=\sqrt{\frac{1}{N}}\sum_{\ell}e^{-ik^{}_{n}
\ell}
c^{\dagger}_{\ell}\ ,\nonumber\\
  k^{}_{n}&=\frac{2\pi n}{N}\ ,\ \ n=0,1,\ldots,N-1\ .
\end{align}
We can write the free particle Hamiltonian as
\begin{eqnarray}
H_o &=& \sum_n \epsilon^o_{k_n}
c^{\dagger}_{k^{}_{n}}c^{}_{k^{}_{n}}
\nonumber \\
 &=& 2\Delta_o \sum_n \cos (k^{}_{n}-\Phi/N)c^{\dagger}_{k^{}_{n}}c^{}_{k^{}_{n}}\
 \label{eq:H_phi}
\end{eqnarray}
Then in this basis we can write:
\begin{equation}
V_{int} = \sum_{k}^{N_s} \sum_n \left[  \mbox{\boldmath
$F$}^k_{n}(\mbox{\boldmath $\sigma$}_k) \rho({k_n}) \;+\;
\mbox{\boldmath $G$}_{n}^k(\mbox{\boldmath $\sigma$}_k)
\hat{c}_{k_n}^{\dagger}\hat{c}_{k_n} \right]
 \label{spin-z3}
\end{equation}
where $\rho({k_n}) = \sum_{n'}\hat{c}_{k_n
+k_{n'}}^{\dagger}\hat{c}_{k_{n'}}$ is the density operator in
momentum space for the particle, and the Fourier-transformed
interaction functions are
\begin{eqnarray}
\mbox{\boldmath $G$}_{n}^k(\mbox{\boldmath $\sigma$}_k) &=&
\sum_{ij} e^{i k_n(i-j)}\mbox{\boldmath $G$}^k_{ij}(\mbox{\boldmath
$\sigma$}_k)
\nonumber\\
\mbox{\boldmath $F$}_{n}^k(\mbox{\boldmath $\sigma$}_k)
  &=& \sum_j e^{i {k_n} j} \mbox{\boldmath
$F$}^k_j(\mbox{\boldmath $\sigma$}_k)
 \label{FG-p}
\end{eqnarray}
In this basis the band Hamiltonian $H_{band}$ has a dispersion which
is a functional of the bath spin distribution:
\begin{align}
H_{band} = \sum_k  \sum_{n} &\epsilon_{k_n} [\mbox{\boldmath
$\sigma$}_k]\hat{c}_{k_n}^{\dagger}\hat{c}_{k_n} \nonumber\\
 &+\;
\sum_{n, n'} v_{n}[\mbox{\boldmath $\sigma$}_k] \hat{c}_{k_n
 + k_{n'}}^{\dagger}\hat{c}_{k_{n'}}
 \label{H-band}
\end{align}
and in which the 'band energy' $\epsilon_{k_n} [\mbox{\boldmath
$\sigma$}_k]$ and the 'scattering potential' $v_{n}[\mbox{\boldmath
$\sigma$}_k]$ are now both functionals over the spin bath
coordinates $\{ \mbox{\boldmath $\sigma$}_k \}$:
\begin{eqnarray}
\epsilon_{k_n}[\mbox{\boldmath $\sigma$}_k] &=& \epsilon_{k_n}^o +
\sum_k \mbox{\boldmath $G$}_{n}^k(\mbox{\boldmath $\sigma$}_k)
\nonumber\\
v_{n}[\mbox{\boldmath $\sigma$}_k] &=& \sum_k \mbox{\boldmath
$F$}_{n}^k(\mbox{\boldmath $\sigma$}_k)
 \label{disp2}
\end{eqnarray}

Under many circumstances one can assume that this symmetry under
rotations also applies to the bath couplings, so these no longer
depend on site variables, ie., $\mbox{\boldmath $F$}_j^k \rightarrow
\mbox{\boldmath $F$}^k$, and $\mbox{\boldmath $G$}_{ij}^k
\rightarrow \mbox{\boldmath $G$}^k$. The results then simplify a
great deal; $\mbox{\boldmath $G$}_{n}^k(\mbox{\boldmath $\sigma$}_k)
\rightarrow 2\mbox{\boldmath $G$}^k(\mbox{\boldmath $\sigma$}_k)
\cos k_n$, and $\mbox{\boldmath $F$}_{n}^k \rightarrow
\mbox{\boldmath $F$}^k$.

Now let us consider the microscopic origin of this model (ie.,
before truncation to the lowest band). The most obvious interaction
between the particle moving around the ring and a set of bath spins
has the local form \cite{dube98}:
 \begin{align}
 H_{int}({\bf R}, \mbox{\boldmath $\sigma$}_k)\;=\; & \sum_k
 \mbox{\boldmath $F$} (\mbox{\boldmath $R$}-\mbox{\boldmath
 $r$}_k)\cdot\mbox{\boldmath $\sigma$}_k
 \nonumber \\
 \;\equiv\; &\sum_k
H_{int}^k({\bf R}, \mbox{\boldmath $\sigma$}_k) \label{eq:H_int}
 \end{align}
where $\mbox{\boldmath $F$} (\mbox{\boldmath $r$})$ is some vector
function, and $\mbox{\boldmath $r$}_k$ is the position at the $k$-th
bath spin. The diagonal coupling $\mbox{\boldmath $F$}_j^k$, or its
linearized form $\mbox{\boldmath $\gamma$}_{k}^j $, is then easily
obtained from \eqref{eq:H_int} when we truncate to the single band
form. But the term (\ref{eq:H_int}) must also generate a
non-diagonal term, which is more subtle. We can see this by defining
the operator
 \begin{equation}
 \hat{T}_{ij}^k\;=\;
\exp{[-i/\hbar\;\int_{\tau_{in}({\bf R}_i)}^{\tau_f({\bf
R}_j)}d\tau\;H_{int}^k({\bf R}, \mbox{\boldmath $\sigma$}_k)]}
 \end{equation}
where the particle is assumed to start in the $i$-th potential well
centered at position ${\bf R}_i$, at the initial time $\tau_{in}$,
and finish at position ${\bf R}_j$ in the adjacent $j$-th well at
time $\tau_f$; the intervening trajectory is the instanton
trajectory (which in general is modified somewhat by the coupling to
the spin bath). Now we operate on $\mbox{\boldmath $\sigma$}_k$ with
$\hat{T}_{ij}^k$, to get
\begin{equation}
|\mbox{\boldmath $\sigma$}^f_k\rangle\;=\;
\hat{T}_{ij}^k\;|\mbox{\boldmath $\sigma$}^{in}_k\rangle\;=\;e^{i
(\phi_k^{ij} + \mbox{\boldmath $\alpha$}_k^{ij}\cdot\mbox{\boldmath
$\sigma$}_k)}|\mbox{\boldmath $\sigma$}^{in}_k\rangle
\end{equation}
where we note that both the phase $\phi_k^{ij}$ multiplying the unit
Pauli matrix $\sigma_k^0$, and the vector
$\mbox{\boldmath$\alpha$}_k^{ij}$ multiplying the other 3 Pauli
matrices $\sigma_k^x, \sigma_k^y, \sigma_k^z$, are in general
complex. In this way the instanton trajectory of the particle acts
as an operator in the Hilbert space of the $k$-th bath spin
\cite{PS00,PS93}. Note that one important implication of this
derivation is that typically $|\mbox{\boldmath$\alpha$}_k^{ij}| \ll
1$, in fact exponentially small, since the interaction energy scale
set by $|\mbox{\boldmath $F$} (\mbox{\boldmath $R$}-\mbox{\boldmath
$r$}_k)|$ is usually much smaller than the "bounce energy" scale
$\hbar\Omega_o$ set by the potential $U({\bf R})$, ie., the
tunneling of the particle between wells is a sudden perturbation on
the bath spins \cite{PS00}. Detailed calculations in specific cases
\cite{PS93,PS00,TPS96} show that $|\mbox{\boldmath$\alpha$}_k^{ij}|
\sim \pi |\mbox{\boldmath $\omega$}_{k}^{ij}|/2\Omega_o$ in this
'sudden' regime, where $\mbox{\boldmath $\omega$}_{k}^{ij} =
\mbox{\boldmath $\gamma$}_{k}^{j} - \mbox{\boldmath
$\gamma$}_{k}^{i}$ is the change in the diagonal coupling acting
between the particle and the $k$-th bath spin when the particle hops
from site $i$ to site $j$ (this result can be found directly from
time-dependent perturbation theory in the sudden approximation).

From these considerations we see that, starting from a ring with the
particle-bath interaction given in (\ref{eq:H_int}), we will end up
with an effective Hamiltonian for the lowest band of the form given
in (\ref{H-band1}), in which the non-diagonal interaction
$\mbox{\boldmath $G$}_{ij}^k(\mbox{\boldmath $\sigma$}_k)$ in
(\ref{spin-z}) has assumed a rather special form.

One can in fact have a more general form for $\mbox{\boldmath
$G$}_{ij}^k(\mbox{\boldmath $\sigma$}_k)$ in the lowest-band
approximation, provided one also introduces in the microscopic
Hamiltonian a coupling
 \begin{equation}
 H_{int}({\bf P}, \mbox{\boldmath $\sigma$}_k)\;=\;\sum_k
\mbox{\boldmath $G$} (\mbox{\boldmath $P$}, \mbox{\boldmath
$\sigma$}_k)
 \label{eq:H_int2}
 \end{equation}
to the momentum of the particle. This can include various terms,
including functions of $\mbox{\boldmath $P$} \times \mbox{\boldmath
$\sigma$}_k$ and $\mbox{\boldmath $P$} \cdot \mbox{\boldmath
$\sigma$}_k$; a detailed analysis is fairly lengthy. The main new
effect of these is to generate terms in the band Hamiltonian which
couple the spins to the amplitude of $t_{ij}$ as well as to its
phase; these do not appear in (\ref{H-band1}).

In any case, if we know $U(\mathbf{R})$, $\mbox{\boldmath $F$}
(\mbox{\boldmath $R$}-\mbox{\boldmath $r$}_k)$, and $\mbox{\boldmath
$G$} (\mbox{\boldmath $P$}, \mbox{\boldmath $\sigma$}_k)$, we can
clearly then calculate all the parameters in the generic model
Hamiltonian, using various methods \cite{PS00,TPS96}. However we are
not interested here in the generic case, since our main object is to
study the dynamics of decoherence in a ring model which contains
only phase decoherence. We therefore make the following
approximations:

(i) We drop the interaction $V_{kk'}^{\alpha \beta}$, between bath
spins (often a very good approximation, since interactions between
defects or nuclear spins are often very weak), and also neglect the
local fields $\mbox{\boldmath $h$}_k$ acting on the
$\{\mbox{\boldmath $\sigma$}_k \}$. Thus we make $H_{SB} = 0$.

(ii) We drop the momentum coupling $\mbox{\boldmath
$G$}(\mbox{\boldmath $P$}, \mbox{\boldmath $\sigma$}_k)$ entirely,
and in the band Hamiltonian (\ref{H-band1}) we drop the diagonal
interaction $\mbox{\boldmath $\gamma$}_k^{j}$. This implies that the
energy of the $k$-th bath spin does not depend on whether the $j$-th
site is occupied. We make this approximation (in many cases not
physically reasonable) only because we wish to study phase
decoherence without the complication of energy relaxation.

(iii) We assume a symmetric ring, so that $\varepsilon_j \rightarrow
0$ and $t_{ij} \rightarrow \Delta_o$ as before; and we absorb the
phases $\phi_k^{ij} \rightarrow \phi_k$ into a renormalization of
$\Delta_o$ (from $\sum_k \mbox{Im}\;\phi_k$), and of $A^o_{ij}$
(from $\sum_k \mbox{Re}\;\phi_k$).

The resulting model $H_{\phi}$ is then just that given in
\eqref{eq:H_phi'}. This turns out to be explicitly solvable, and
reveals some important properties of phase decoherence. We will
usually assume the parameters $\mbox{\boldmath$\alpha$}_k^{ij}$ are
small, in line with the remarks above (although the net effect of
all of them may be very large), and we will also usually specialize
to the case $\mbox{\boldmath$\alpha$}_k^{ij} \rightarrow
\mbox{\boldmath$\alpha$}_k$, consistent with a completely symmetric
ring.

Finally, let us briefly compare with the kind of Hamiltonian one
would expect for a particle on a ring coupled to an oscillator bath.
Let us assume a set of oscillators with Hamiltonian $H_o + H_{osc} +
H_{int}$, where $H_o$ is again the free particle hopping
Hamiltonian, coupling to a set of $N_o$ oscillators with Hamiltonian
\begin{equation}
H_{osc} =  \sum_{q = 1}^{N_o} {1 \over 2}( {p_q^2 \over m_q} + m_q \omega_q^2 x_q^2 )  \\
 \label{H-osc}
\end{equation}
In general there will be diagonal couplings $\{ V_j(q) \}$ and
non-diagonal couplings $\{ U_{ij}(q) \}$ between particle and
oscillators. We could also have a coupling to the oscillator momenta
- however in this case one can make a canonical
transformation\cite{ajl84} which transforms this back into a
coupling to the $\{ x_q \}$. Typically the couplings $\{ V_j(q)
,U_{ij}(q) \} \sim O(N_o^{-1/2})$. We note here that in many
microscopic models of this kind, the couplings $\{ V_j(q) ,U_{ij}(q)
\}$ are actually also strong functions of temperature, either
because the underlying effective Hamiltonian is strongly
$T$-dependent (eg., in a superconductor \cite{amb83}), or because
the coupling to the oscillators is non-linear (eg., in the coupling
to a soliton \cite{wada78}).

If we restrict the problem to rotationally invariant couplings on
the ring, then we can write
\begin{equation}
H_{int} = \sum_q \left[ \sum_{<ij>} (U_q
\hat{c}_{i}^{\dagger}\hat{c}_j + H.c.)  + \sum_j V_q
\hat{c}_{j}^{\dagger}\hat{c}_j \right] x_q
 \label{V-trans}
\end{equation}
where $U_{ij}(q) \rightarrow U_q$, $V_j(q) \rightarrow V_q$, and the
sum $\sum_{<ij>}$ is over nearest neighbours. It is then
straightforward to go through the same manipulations as in
(\ref{spin-z3})-(\ref{disp2}), to get a renormalised band which is a
functional of the $\{ x_q \}$.

In these results there is no connection between the ring sites and
the space in which the oscillators are supposed to exist. However in
many cases the oscillator displacement field $x_j$ can be defined at
each site $j$ of the ring; the coupling then reduces to
\begin{align}
\hat{V} \;=\;& \sum_{\bf q}\sum_j v_{\bf q}
\hat{c}_{j}^{\dagger}\hat{c}_je^{i{\bf q}\cdot {\bf R}_j}x_{\bf q}
 \nonumber \\
\equiv \;& \sum_{jj'} v({\bf R}_j - {\bf r}_{j'})
\hat{c}_{j}^{\dagger}\hat{c}_j x({\bf r}_{j'})
 \label{V-osc}
\end{align}
in which ${\bf R}_j$, ${\bf r}_{j'}$ are site vectors on the ring,
and $x_{\bf q}$ is now the Fourier transform of $x_j$.


\section{ Free band particle Dynamics}
 \label{sec:bareR}


We first consider the dynamics of a free particle in some initial
state moving on the symmetric $N$-site ring described by $H_o$ in
\eqref{eq:H_phi}, with no bath.

\begin{figure}[h]
\includegraphics[width=7cm]{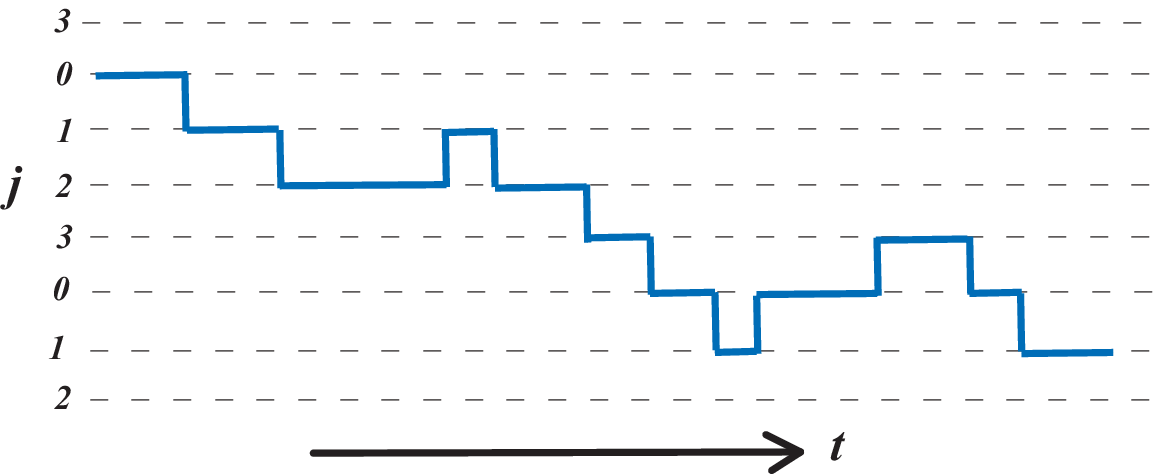}
\caption{\label{Fig:PATHS} Colour online: A particular path in a
path integral for the particle, shown here for an $N=3$ ring. This
path, from site $0$ to site $1$, has winding number $p=1$.}
\end{figure}

For this free particle the dynamics is entirely described in terms
of the bare 1-particle Green function
\begin{align}
 G^o_{jj'}(t) &\equiv \langle j|G^o(t)|j'\rangle
\equiv\langle j|e^{-i{\cal H}_ot}|j'\rangle\nonumber\\
&=\frac{1}{N}\sum_{n}e^{-i2\Delta^{}_{0}t\cos (k^{}_{n}-\Phi/N
)}e^{ik^{}_{n}(j' -j)}\ ,
 \label{GLM}
\end{align}
which gives the amplitude for the particle to propagate from site
$j'$ at time zero to site $j$ at time $t$. These paths are rather
simple (see Fig. \ref{Fig:PATHS}); they can be labelled by the
initial and final sites, and by the winding number of the path
around the ring.

The 1-particle Green function can be evaluated in various ways (see
Appendix); the result can be usefully written as
\begin{equation}
G^o_{jj'}(t)=\sum_{p=-\infty}^{+\infty}J_{Np+j'-j}(2\Delta_o t)
e^{-i (Np+j'-j) (\Phi/N+\pi/2)}
 \label{GJJ}
\end{equation}
where $\sum_p$ is a sum over winding numbers. The "return amplitude"
$G^o_{00}(t)$ is then given by
\begin{align}
G^{o}_{00}(t)&=\sum_{p}e^{-ip\Phi}(-i)^{|Np|}J_{|Np|}(2\Delta^{}_{o}t)
\nonumber \\
&=\sum_{p}e^{-ip\Phi}I_{Np}(-2i\Delta^{}_{o}t)
 \label{G00}
\end{align}
where in the last form we use the hyperbolic Bessel function.

It is often more useful to have expressions for the density matrix;
even though these depend trivially for a free particle on the Green
function, they are essential when we come to compare with the
reduced density matrix for the particle coupled to the bath. One
has, for the 'bare' density matrix operator of the system at time
$t$,
\begin{align}
\hat{\rho}^o(t)=e^{-i{\cal H}_ot}\hat{\rho}^o(0)e^{i{\cal H}_ot}.
\end{align}
Thus, suppose we have an initial density matrix  $\rho^{(in)}_{ll'}
= \langle l |\rho(t=0)|l' \rangle$ (where $l$ and $l'$ are site
indices), then at a later time $t$ we have
\begin{align}
\rho^o_{jj'}(t) \equiv \langle j|\hat{\rho}^o(t)|j'\rangle &=\langle
j|e^{-i{\cal
H}_ot}|l\rangle  \rho_{ll'}^{(in)} \langle l' |e^{i{\cal H}_ot}|j'\rangle \nonumber\\
 &=\rho_{ll'}^{(in)} G^o_{jl}(t) G^o_{j'l'}(t)^{\dagger},
 \label{rho-ot}
\end{align}
where we use the Einstein summation convention (summing over
$l,l'$). This equation defines the propagator $K^o_{jj',ll'}(t)$ for
the free particle density matrix, as
\begin{equation}
K^o_{jj',ll'}(t) =  G^o_{jl}(t)
G^o_{j'l'}(t)^{\dagger}.\label{rho-ot2b1}
\end{equation}

In the main text of this paper we will almost always quote results
for the special case where the particle begins at $t=0$ on site $0$.
In the case of the free particle, this means that $\rho_{ll'}^{(in)}
= \delta_{0 l} \delta_{l' 0}$, and only the propagator matrix
$K^o_{jj',00}(t)$ enters the results; then we have
\begin{equation}
\langle j|\hat{\rho}^o(t)|j'\rangle \rightarrow K^o_{jj',00}(t) =
G^o_{j0}(t) G^o_{j'0}(t)^{\dagger}.
\end{equation}
In the Appendix we give the results for an arbitrary initial density
matrix.

The evaluation of the time-dependent density matrix for the free
particle turns out to be quite interesting mathematically. As
discussed in the Appendix, one can evaluate $\rho^o_{jj'}(t)$ as a
sum over pairs of paths in a path integral, to give a double sum
over winding numbers, or else as a single sum over winding numbers.
Consider first the double sum form; again, for the special case
where $\rho^{(in)}_{ll'} = \delta^{}_{0 l} \delta^{}_{l' 0}$ (the
particle starts at the origin), this can be written as
\begin{widetext}
\begin{equation}
\rho^{o}_{jj'}(t)=\sum_{pp'}e^{i(p-p')\Phi}e^{i\Phi
(j-j')/N}(-i)^{Np +j}(i)^{Np'+j'}
 J^{}_{Np+j}(2\Delta^{}_{o}t)J^{}_{Np'+j'}(2\Delta^{}_{o}t) ,
 \label{rho-o1}
\end{equation}
\end{widetext}
where $p,p'$ are the winding numbers (see Appendix for the
derivation for a general initial density matrix). This form has a
simple physical interpretation - the particle propagates along pairs
of paths in the density matrix, one finishing at site $j$ and the
other at site $j'$, and the order of each Bessel function simply
gives the total number of sites traversed in each path, with
appropriate Aharonov-Bohm phase multipliers for each path.

If one instead writes the answer as a single sum over winding
numbers, again assuming $\rho^{(in)}_{ll'} = \delta^{}_{0 l}
\delta^{}_{l' 0}$, we get:
\begin{widetext}
\begin{equation}
\rho^o_{jj'}(t)=\frac{1}{N}\sum_{m=0}^{N-1} \sum_{p'=-\infty}^\infty
J^{}_{Np'+j'-j}[4\Delta^{}_o t \sin
(k^{}_m/2)]e^{i\Phi[p'+(j'-j)/N]-ik^{}_m(j+j'-Np')/2}\
 \label{rho1}
\end{equation}
\end{widetext}
where as before the $\{ k_m \}$ are the momenta of the particle
eigenfunctions. The physical interpretation of this form is less
obvious, but the sums are much easier to evaluate since they only
contain single Bessel functions instead of pairs of them. Thus
wherever possible we reduce double sum forms to single sums. Notice
that for these finite rings, the bare density matrix is of course
strictly periodic in time. Notice also that the diagonal elements of
$\rho(t)$ are generally periodic in $\Phi$. However, the
off-diagonal elements are only periodic in $\Phi/N$. In contrast,
$e^{i\Phi(j-j')/N}\langle j|\rho(t)|j'\rangle$ is periodic in
$\Phi$, with period $2\pi$. This latter is the quantity needed for
calculating the currents, as we will see below.

From either $G^{o}_{jj'}(t)$ or $\rho^{o}_{jj'}(t)$ we may
immediately compute two useful physical quantities. First, the
probability $P^o_{j0}(t)$ to find the particle at time $t$ at site
$j$, assuming it starts at the origin; and second, the current
$I^o_{j,j+1}(t)$ between adjacent sites as a function of time.

Looking first at the probability $P^o_{j0}(t)$, one has
\begin{align}
P^{o}_{j0}(t)=\langle j | \hat{\rho}^o(t)|j\rangle = |G^o_{j0}(t)|^2
\end{align}
which from above can be written in double sum form as
\begin{align}
P^{o}_{j0}(t)=&\sum_{pp'}J_{Np+j}(2\Delta_o t) J_{Np'+j}(2\Delta_o
t) \nonumber\\
 & \;\;\;\;\;\;\;\;\; \times \; e^{-iN(p'-p)(\Phi/N+\frac{\pi}{2})}
 \label{Pjo-free}
\end{align}
or in single sum form as
\begin{align}
P^{o}_{j0}(t)=&\frac{1}{N}\sum_{m=0}^{N-1} \sum_{p=-\infty}^\infty
e^{ip(\Phi+Nk_m/2)} \nonumber\\
 \;\;\;\;\;\;\;\;\;\;\;\;\;\;\;\;\;\;& \times
J_{Np}[4\Delta_o t \sin (k^{}_m/2)]\ .
 \label{rho1A}
\end{align}


One may also compute moments of these probabilities. These are not
terribly meaningful for a small ring, because any wave-packet will
be spread around the ring. However for a large ring they can be
useful- for example, the 2nd moment $\sum_j j^2P^o_{j0}(t)$ tells us
the rate at which an initial density matrix spreads in time,
provided the spatial extent of the density matrix is much smaller
than the ring circumference. Coherent dynamics will then manifest
itself as ballistic propagation of an initial wave-packet.

\begin{figure}[h]
\includegraphics[width=4cm]{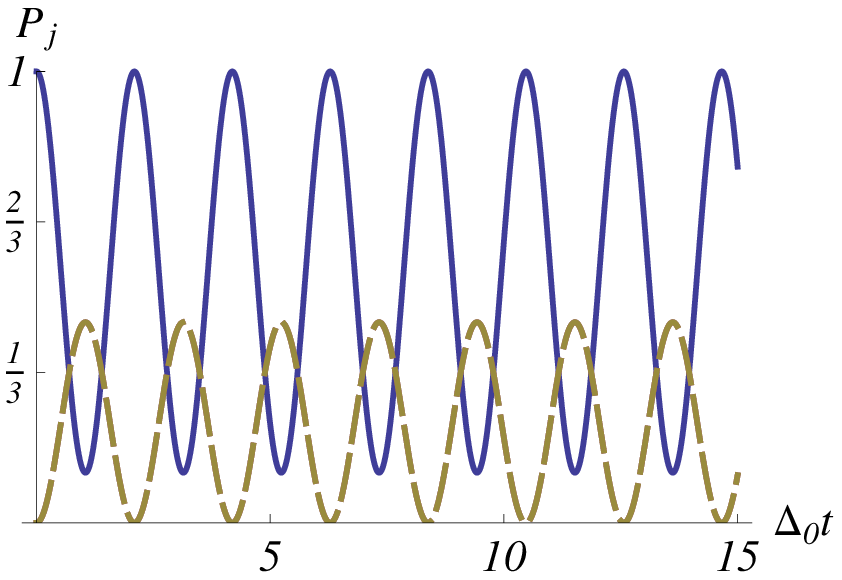}\ \ \
\includegraphics[width=4cm]{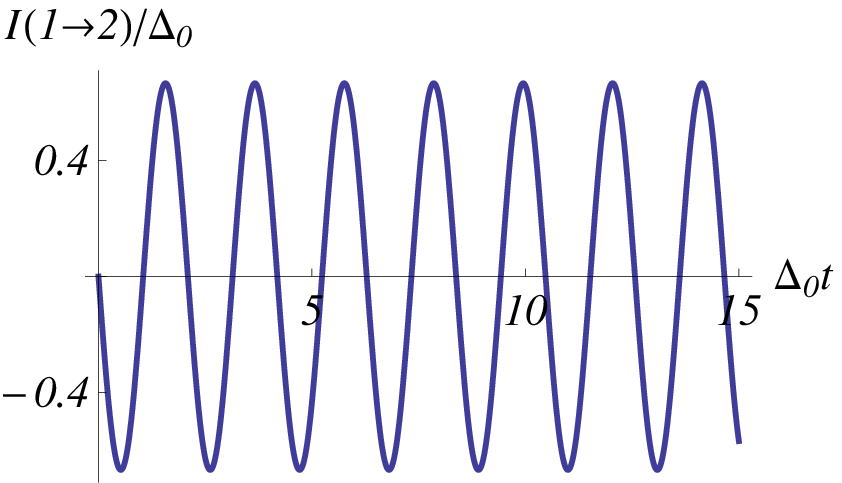}\\
\includegraphics[width=4cm]{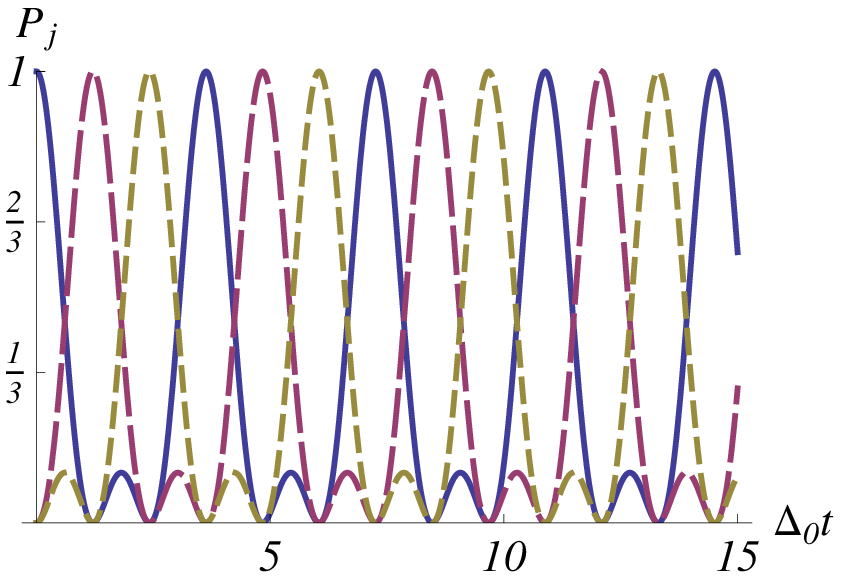}\ \ \
\includegraphics[width=4cm]{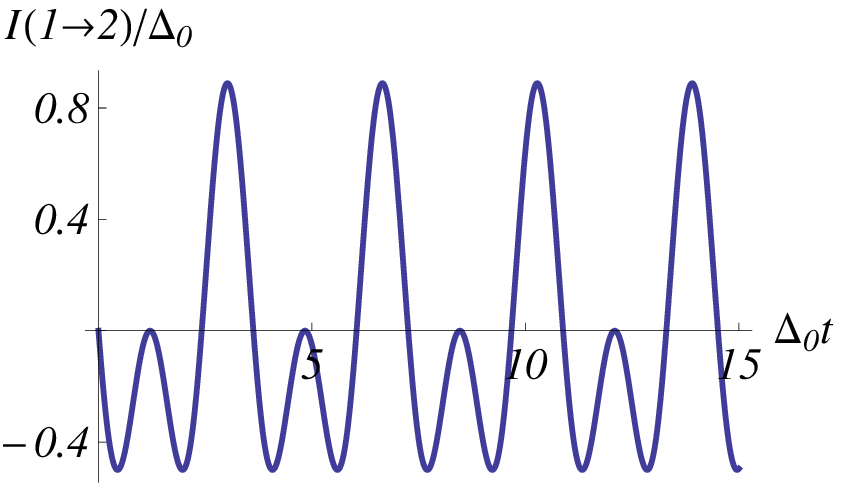}\\
\caption{Colour online: Results for the free particle for $N=3$ and
for a particle initially on site $1$. Left: The probabilities to
occupy site $1$ (full line), $2$ (large dashes), and $3$ (small
dashes). Right: the current from site $1$ to site $2$. Top:
$\Phi=0$. Bottom: $\Phi=\pi/2$. } \label{1}
\end{figure}

From these general expressions it is hard to see what is going on.
To give some idea of how the probability density behaves, it is
useful to then look at these results for a small 3-site ring, where
the oscillation periods are quite short. One then has, for the case
where the particle starts at the origin, that
\begin{align}
&P^{o}_{j0}(t)=\frac{1}{3}\bigl (1+(3\delta^{}_{j,0}-1)\bigl
[J^{}_0(2\Delta_o\sqrt{3}t)\nonumber\\
&+2\sum_{p=1}^\infty
J^{}_{6p}(2\Delta_o\sqrt{3}t)\cos(2p\Phi)\bigr ] \nonumber\\
&+(\delta^{}_{j,1}-\delta^{}_{j,2})2\sqrt{3}\sum_{p=1}^\infty
J^{}_{6p-3}(2\Delta_o\sqrt{3}t)\sin((2p-1)\Phi)\bigl )\ .\label{Pj}
\end{align}
In Fig. \ref{1} the return probability $P^{o}_{00}(t)$ is plotted
for the case $N=3$, using (\ref{Pj}). From the results one striking
feature immediately emerges - we see that the periodic behaviour
depends strongly on the flux $\Phi$. This flux dependence
illustrates the way in which the flux controls the particle
dynamics, by acting directly on the particle phase. In section V we
will see how this also happens when one looks at interference
between 2 wave-packets; and in sections IV and V we will see how
decoherence washes out the flux dependence of the particle dynamics.
Thus the flux dependence of the particle dynamics very effectively
measures how coherent its dynamics may be.

Turning now to the current $I^{o}_{j,j+1}(t)$ from site $j$ and
site $j+1$, this is given from elementary quantum mechanics by
\begin{align}
I^{o}_{j,j+1}&(t) \;\;=\;\; 2\; {\rm Im} \; [\Delta_o e^{-i\Phi/N}
\rho^o_{j,j+1}(t)] \nonumber\\
&= \;\; i\Delta^{}_{o}\Bigl
(e^{i\Phi/N}\rho^{o}_{j+1,j}(t)-e^{-i\Phi/N}\rho^{o}_{j,j+1}(t)\Bigr
)
 \label{PER}
\end{align}
where the flux per link appears in each contribution. Again, one can
write this expression as either a double sum over pairs of winding
numbers, or as a single sum (see Appendix for the general results
and derivation). For the case where the particle starts from the
origin, these expressions reduce to
\begin{widetext}
\begin{align}\label{eq:I_nn}
I^o_{j+1,j}\;=&\; 2\Delta_o\sum_{pp'} J_{Np+j}(2\Delta_o t)
J_{Np'+j+1}(2\Delta_o t) \cos[(\frac{\pi}{2}N+\Phi)(p'-p)]
\nonumber\\
=&\; \frac{2\Delta_o}{N}\sum_{m=0}^{N-1}\sum_p J_{Np+1}(4\Delta_o
t\sin\frac{k_m}{2})e^{-ik_m(\frac{Np+1}{2}+j)}i^{Np+1}\cos[(\frac{\pi}{2}N+\Phi)p]
\end{align}
\end{widetext}
for the double and single sum forms respectively.

Again, the currents across any links must be strictly periodic in
time for this free particle system; and again, it is useful to show
the results for a 3-site system. For this case $N=3$, and assuming
that the particle begins at the origin, we find
\begin{equation}
\begin{split}
I^o_{0,1}\; = \;\frac{2\Delta_o}{3}& {\rm Im} \sum_{m=1}^{2}\sum_p
J_{3p+1}(4\Delta_o t\sin\frac{m\pi}{3}) \\
&\times e^{-im\pi(3p+1)/3}i^{3p+1}\cos[(\frac{3\pi}{2}+\Phi)p]
\end{split}
\end{equation}
which we can also write in the form
\begin{align}
I^o_{0,1}\;=\; &\frac{2\Delta_o}{3} {\rm Im} \sum_p
J_{3p+1}(2\sqrt{3}\Delta_o t)
\cos[(\frac{3\pi}{2}+\Phi)p] \nonumber\\
& \times
i^{3p+1}\sum_{m=1}^{2}(e^{-i\pi(3p+1)/3}+e^{-i2\pi(3p+1)/3})
\end{align}
Now let us write
$(e^{-i\pi(3p+1)/3}+e^{-i2\pi(3p+1)/3})=(-1)^{p}e^{-i\pi/3}+e^{-2i\pi/3}$.
If $p$ is even, this becomes $-i\sqrt{3}$ and
$\cos[(\frac{3\pi}{2}+\Phi)p]=(-1)^{3p/2}\cos(\Phi p)$; If $p$ is
odd, it becomes $-1$ and
$\cos[(\frac{3\pi}{2}+\Phi)p]=(-1)^{3(p-1)/2}\sin(\Phi p)$.
Therefore, we have
\begin{align}
I^o_{0,1}=&\frac{2}{3}\Delta_o\sum_{p=-\infty}^\infty
J^{}_{3p+1}(2\Delta_o\sqrt{3}t)K(p,\Phi)\
,\nonumber\\
&K(p,\Phi)=\sin(p\Phi)\ \ \ {\rm if}\ \ p={\rm odd}\ ,\nonumber\\
&K(p,\Phi)=\sqrt{3}\cos(p\Phi)\ \ \ {\rm if}\ \ p={\rm even}\
.\label{I12}
\end{align}
These results are also shown in Fig. \ref{1}.  Notice that in this
special case the result is periodic in $\Phi$; this is not however
true for a general initial density matrix $\rho^{(in)}_{ll'}$, when
the result is periodic in $\Phi/N$.



\section{ Ring plus Bath: Phase Averaging}
\label{sec:R+Bath}


We now wish to solve for the dynamics of the particle once it is
coupled to the bath, via the Hamiltonian \eqref{eq:H_phi'}. This is
done in general by integrating out the bath spins, to produce
expressions for the reduced density matrix of the particle. In this
section we first show how this is done, and then give results for
physical quantities (in particular, the probability $P_{j0}(t)$ and
the current $I_{j,j+1}(t)$). Finally, we briefly compare the results
to the behaviour one expects for a ring coupled to an oscillator
bath.


\subsection{General results} \label{sec:genRes-B}


As shown in the Appendix, the reduced density matrix for the
particle obeys the equation of motion
\begin{equation}
\rho_{jj'}(t) = \sum_{l,l'} {\cal K}_{jj',ll'}(t) \rho^{(in)}_{l,l'}
\label{rho-ot2''}
\end{equation}
where ${\cal K}_{jj',ll'}(t)$ is the propagator for the reduced
density matrix. This latter can be written in the form of a double
sum over winding numbers
\begin{equation}
{\cal K}_{jj',ll'}(t) =  \sum_{pp'} K^o_{jj',ll'}(p,p';t)
F_{jj'}^{ll'}(p,p')
 \label{K-F}
\end{equation}
where the function $K^o_{jj',ll'}(p,p';t)$ is the free particle
propagator for fixed winding numbers $p,p'$ (so that
$K^o_{jj',ll'}(t) = \sum_{pp'} K^o_{jj',ll'}(p,p';t)$; see the
Appendix, eqtn. (\ref{Ko-Kopp'}) {\it et seq.}). All effects from
the spin bath are then contained in $F_{jj'}^{ll'}(p,p')$, which we
will call the "influence function". The remarkable thing is that
this function depends only on the initial and final states, and on
the winding numbers - all other aspects of the two paths involved in
the density matrix propagation have disappeared. As explained in the
appendix, this is a particular feature of the pure phase decoherence
being treated here.

The form the influence function takes depends on what kind of
averaging we do over the bath. To discuss this, let us first
discriminate between two different ways of averaging over the bath,
as follows:

(i) The first and most obvious case is where the $\mbox{\boldmath
$\alpha$}_k^{mn}$ are considered to be a set of fixed couplings, for
a specific single ring. In this case the average is only over the
bath states; we will denote this bath average by $<....>$. Often it
will only involve a thermal average over the bath states.

(ii) However it is often the case that one is either interested in
an ensemble of rings, all having the same free particle Hamiltonian
but with the $\mbox{\boldmath $\alpha$}_k$ possibly varying from one
ring to another, or a single ring in which the values of the
couplings $\mbox{\boldmath $\alpha$}_k$ are indeterminate. In this
case it makes sense to define a probability distribution
$P(\mbox{\boldmath $\alpha$})$ over a coupling variable
$\mbox{\boldmath $\alpha$}$. One then must average not only over the
bath states themselves, but also over the bath couplings. We will
denote this double average by $<<.....>>$, to signify the average
over both the bath states and the probability distribution; and the
influence function for this case will be written as
$\Bar{F}_{jj'}^{ll'}(p,p')$, with the bar over the $F$ signifying
that an average over couplings is being done as well.

In general the results for the dynamics of the density matrix and
the current, and their dependence on the influence function, may be
quite complicated. Thus, before we begin quoting results, it is
useful to note what are the important parameters in the problem. We
will only consider here the simplest completely symmetric case where
$ \mbox{\boldmath $\alpha$}_k^{mn} \rightarrow \mbox{\boldmath
$\alpha$}_k$ for all links $\{ mn \}$; and we will assume that $|
\mbox{\boldmath $\alpha$}_k | \ll 1$ for all $k$, as discussed in
section II. Now in the previous literature for this case of pure
phase decoherence, it has been usual to define a 'topological
decoherence' parameter \cite{PS00,PS93}
\begin{equation}
\lambda \;=\; \frac{1}{2}\sum_k |\mbox{\boldmath $\alpha$}_k |^2
 \label{lambda}
\end{equation}
which provides a measure of the strength of the pure phase
decoherence \cite{PS00}. If the number $N_s$ of bath spins is large,
then we can have $\lambda \gg 1$; this is the limit of strong phase
decoherence.

However we shall see in what follows that on a ring it is often more
useful to define a parameter $F_0(\bar{p})$ that also depends on a
winding number $\bar{p}$. The form of this parameter depends on
which of the two bath averages is performed. In the case where only
an average over the bath states is performed, we have
\begin{equation}
F_0(\bar{p}) \;=\; \prod_k \cos(N\bar{p} |\mbox{\boldmath
$\alpha$}_k| )
 \label{F_0'}
\end{equation}
which defines a rather complicated function of the fixed bath
couplings. The strong decoherence limit for this case is defined by
the parameter $\lambda$ defined above.

In the case where we also perform an average over the bath
couplings, we have
\begin{equation}
\bar{F}_0(\bar{p}) = \prod_k\int d \mbox{\boldmath $\alpha$}_k
P(\mbox{\boldmath $\alpha$}_k) \cos(N\bar{p} |\mbox{\boldmath
$\alpha$}_k| )
 \label{F_0-bar}
\end{equation}
The result then depends on what form one has for the distribution
function $P(\mbox{\boldmath $\alpha_k$})$. In what follows we will
use, as an example, a Gaussian distribution, given by
\begin{equation}
P(|\mbox{\boldmath $\alpha$}_k|)  = e^{-|\mbox{\boldmath
$\alpha$}_k|^2/2\lambda^{}_o}/\sqrt{2\pi\lambda^{}_o}
 \label{P-alpha}
\end{equation}
so that
\begin{equation}
\bar{F}_0(\bar{p})=e^{-\lambda N^2\bar{p}^2/2},\ \ \ \
\lambda=N_s\lambda^{}_o\
 \label{lamb-F}
\end{equation}
The limit $\lambda\rightarrow\infty$ is the ``strong decoherence"
limit for this distribution, where we have
$\bar{F}_0(\bar{p})\rightarrow\delta^{}_{\bar{p},0}$. However we
will see below that it is convenient to think of the strong
decoherence regime for the present problem as that for which the
particle dynamics is independent of flux - we will see that this
happens already for quite small values of $\lambda$.

We can see why these functions enter by considering the forms for
$F_{jj'}^{ll'}(p,p')$ and $\bar{F}_{jj'}^{ll'}(p,p')$ that enter
into physical quantities. In the appendix the full expressions for
these are derived; but here we will again only use them for the case
where $\rho^{(in)}_{l,l'} = \delta_{0 l} \delta_{l' 0}$, ie., the
particle starts at the origin, and so only the function
$F_{jj'}(p,p') \equiv F_{jj'}^{00}(p,p')$ comes in. We will also
again assume the purely symmetric case where $\mbox{\boldmath
$\alpha$}_k^{ij} \rightarrow \mbox{\boldmath $\alpha$}_k$ for every
link.

Let us first consider the case of fixed bath couplings. In this case
the form of the influence function reduces to (see Appendix):
\begin{equation}
 \label{eq:F_avg}
F_{jj'}(p,p')\;=\;\langle e^{-i N [(p-p')+(j-j')]\sum_k
\mbox{\boldmath $\alpha$}_k\cdot\mbox{\boldmath $\sigma$}_k} \rangle
\end{equation}
Notice that $F_{jj'}(p,p')$ is a function only of the distance
$j-j'$ between initial and final sites, and of the difference
$\bar{p} = p-p'$ in winding numbers. Writing this now as
$F_{jj'}(\bar{p)}$, let us evaluate it by assuming the usual thermal
initial bath spin distribution. Since all the bath states are
degenerate, then at any finite $T$ all states are equally populated;
we then get:
\begin{equation}
F_{jj'}(\bar{p}) \;=\; \prod_k \cos((N\bar{p}+j-j')|\mbox{\boldmath
$\alpha$}_k|)
 \label{F-prod}
\end{equation}
Other initial non-thermal distributions for the spin bath states are
also easily evaluated from (\ref{eq:F_avg}).


\subsection{Physical Quantities} \label{sec:PhysQ-B}


From expressions like (\ref{F-prod}) one can now write down
expectation values of physical quantities as a function of time. The
simplest example is the probability for the particle to end up at
some site after a time $t$, having started at another. Thus, eg.,
the probability $P_{j0}(t)$ to move to site $j$ from the origin in
time $t$ is now given by
\begin{align}
P_{j0}(t)\;=\;&\rho_{jj}(t) \nonumber\\
=&\; \sum_{pp'}J_{Np+j}(2\Delta_o
t) J_{Np'+j}(2\Delta_o t) \nonumber\\
& \;\;\;\;\times e^{-i N(p'-p)(\Phi/N+\frac{\pi}{2})} \;F_0(p,p')
 \label{Pj0-Bt}
\end{align}
which is a simple generalization of the free particle result in
(\ref{Pjo-free}); we note that only the term
\begin{equation}
F_0(p,p') \;=\; \prod_k \cos(N(p-p') |\mbox{\boldmath $\alpha$}_k| )
 \label{F_0}
\end{equation}
in the influence function survives in this expression. Since this
function depends only on the difference $p-p'$, it is identical to
the function $F_0(\bar{p})$ defined in (\ref{F_0'}) above (letting
$p=\bar{p}$). We shall see below that the ring current is also
controlled by this same function. Note that it has a complex
multiperiodicity, as a function of the $N_s$ different parameters
$N\bar{p}|\mbox{\boldmath $\alpha$}_k|$; we do not have space here
to examine the rich variety of behaviour found in the system
dynamics as we vary these parameters.

Now let us consider the case where we also average over the bath
couplings. One then finds (see appendix) that
\begin{equation}
\bar{F}_{jj'}(\bar{p}) =  \prod_k\int d\mbox{\boldmath $\alpha$}_k
P(\mbox{\boldmath $\alpha$}_k)  \langle e^{-i N [\bar{p}+(j-j')]
\mbox{\boldmath $\alpha$}_k\cdot\mbox{\boldmath $\sigma$}_k} \rangle
 \label{barF-jj'}
\end{equation}
In the symmetric case we can treat each bath spin in the same way,
and simply use a distribution function $P(|\mbox{\boldmath
 $\alpha$}|)$, the same for all the
 different $\{ \mbox{\boldmath $\sigma$}_k \}$. Then we can treat
everything in terms of this single average, over a single
representative spin $\mbox{\boldmath $\sigma$}$ from the bath. Then,
eg., for an initial thermal ensemble for the bath spins, this gives
\begin{equation}
\bar{F}_{jj'}(\bar{p}) =  [\int d\mbox{\boldmath $\alpha$}
P(\mbox{\boldmath $\alpha$}) \cos((N \bar{p}+j-j')|\mbox{\boldmath
$\alpha$}|)]^{N_s}
 \label{barF-jj'T}
\end{equation}

To give something of the flavour of this case, we use the Gaussian
distribution for the $P(|\mbox{\boldmath $\alpha$}|)$, given by
(\ref{P-alpha}) above. Then, for the thermal ensemble just given, we
have
\begin{equation}
\bar{F}_{jj'}(\bar{p}) = \exp [-\lambda(N\bar{p} + j'-j)^2/2]
 \label{barF-G}
\end{equation}
It is then immediately obvious that the result for the probability
for the particle to go from site $0$ to site $j$ in time $t$ is the
same expression as (\ref{Pj0-Bt}) above, but now with
$\bar{F}_0(\bar{p})$ instead of $F_0(\bar{p})$.

\begin{figure}[h]
\includegraphics[width=4cm]{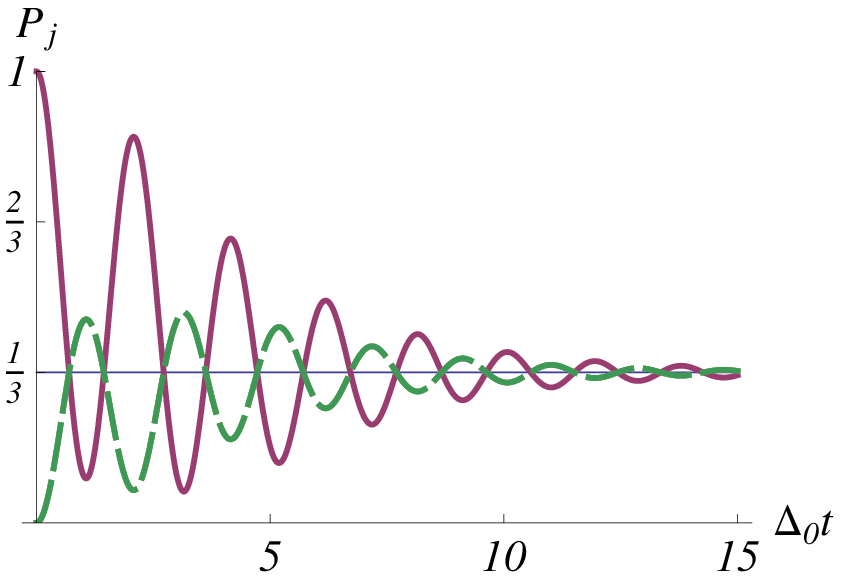}\ \ \
\includegraphics[width=4cm]{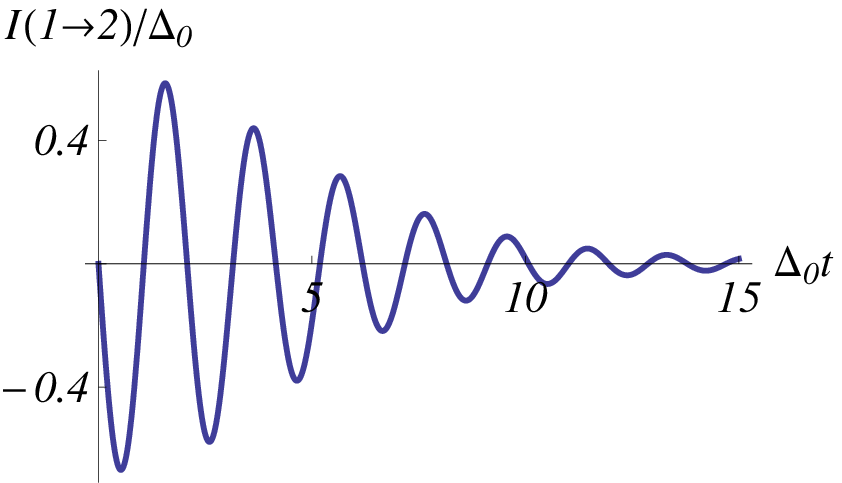}\\
\includegraphics[width=4cm]{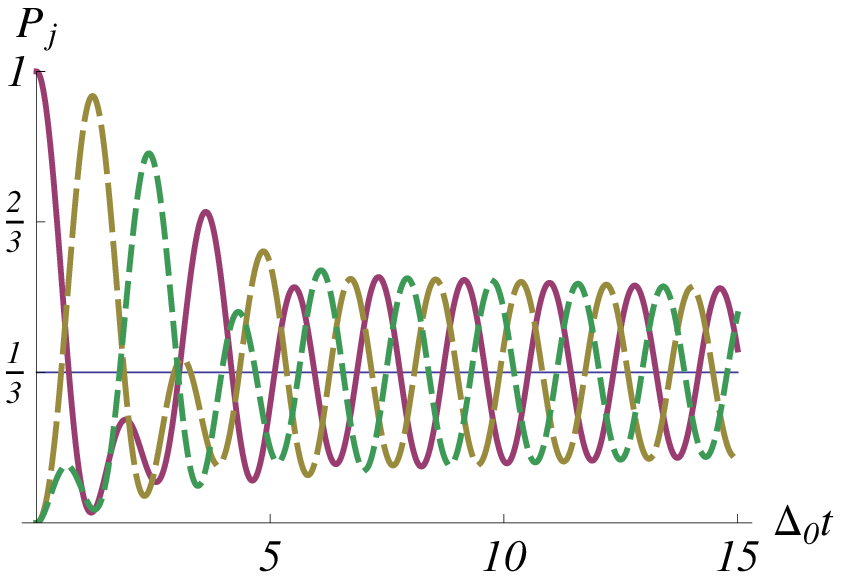}\ \ \
\includegraphics[width=4cm]{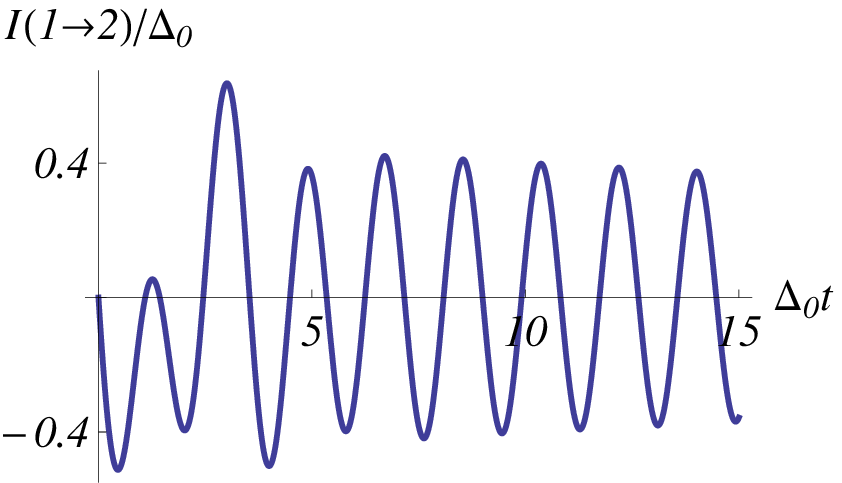}\\
\caption{Colour online: Plot of $P_{j0}(t)$ for a $3$-site ring, for
a particle initially on site $1$, in the intermediate decoherence
limit, with $\lambda=.02$. Left: The probability to occupy site $0$
(full line), $1$ (large dashes), and $2$ (small dashes). Right: the
current from site $0$ to site $1$. Top: $\Phi=0$. Bottom:
$\Phi=\pi/2$ .}
 \label{3}
\end{figure}

To see how this behaves, let us take the specific case where $N=3$
again. Then for this 3-site ring one has, for example, that
\begin{align}
P^{}_{10}(t)=\frac{1}{3}&\bigl (1+2[J^{}_0(2\Delta_o\sqrt{3}t)
\nonumber \\ &+2\sum_{p=1}^\infty J^{}_{6p}(2\Delta_o\sqrt{3}t)
\cos(2p\Phi)\bar{F}_0(6p)]\bigr )\ .
 \label{Pja}
\end{align}
To analyse this result, note that for $x\gg (6p)^2$, we can use
$J^{}_{6p}(x)\approx (-1)^p\sqrt{2/(\pi x)}\cos(x-\pi/4)$. The
function ${\bar F}^{}_0({\bar p})$ decays with ${\bar p}$, and
becomes negligible for large enough ${\bar p}$. For example, Eqtn.
(\ref{lamb-F}) implies that ${\bar F}^{}_0(6p)<e^{-10}$ for ${\bar
p}>p^{}_{max}$,  with $p^{}_{max}=\sqrt{5/9\lambda}$. Neglecting
these terms in the sum in Eq. (\ref{Pja}) we conclude that for
$2\Delta_o\sqrt{3}t\gg (6p^{}_{max})^2$ we have e.g.
\begin{align}
P_{10}(t)&\approx\frac{1}{3}\bigl
[1+\frac{2A}{\sqrt{\pi\Delta_o\sqrt{3}t}}
\cos(2\Delta_o\sqrt{3}t-\pi/4)\bigr ]\ ,\nonumber\\
&A=  1+2 \sum_{p=1}^\infty (-1)^p\cos(2p\Phi)\bar{F}_0(6p)\ .
 \label{Pjaa}
\end{align}
The sum in the amplitude $A$ reduces to $\sum(-1)^p\bar{F}_0(6p)$,
for $\Phi=0$, and to $\sum \bar{F}_0(6p)$, for $\Phi=\pi/2$.
Clearly, switching from $\Phi=0$ to $\Phi=\pi/2$ causes a large
increase in $A$. Notice that the inverse Fourier transform of the
amplitude $A(\phi)$ can be used to measure the decoherence function
$\bar{F}_0(6p)$. Results for this low decoherence regime are shown
in Fig. \ref{3}.

\begin{figure}[h]
\includegraphics[width=4cm]{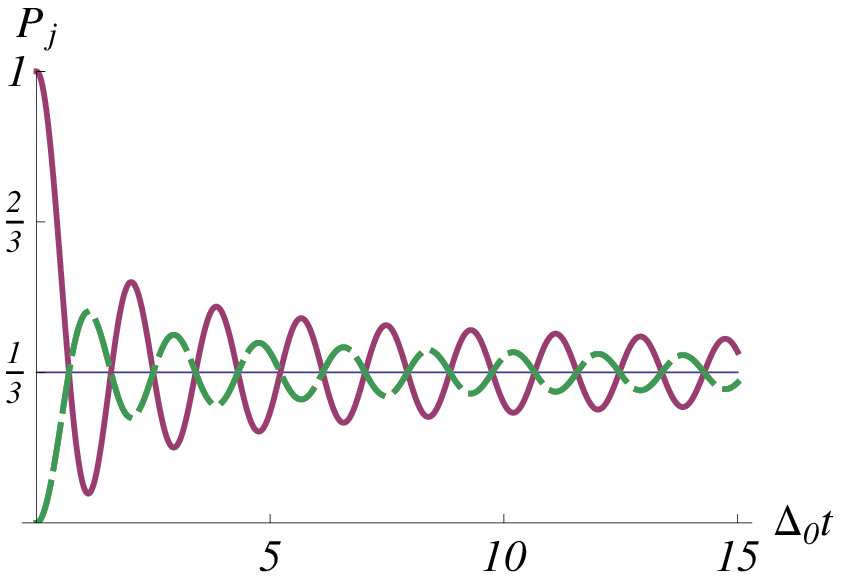}\ \ \
\includegraphics[width=4cm]{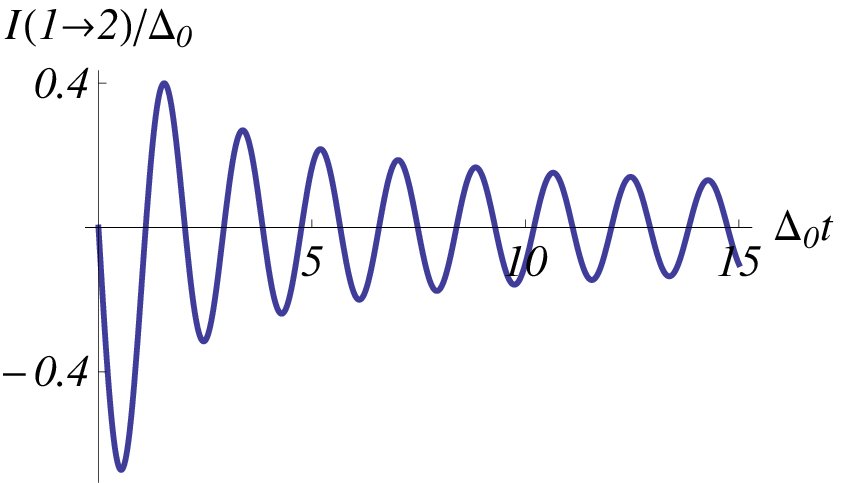}\\
\caption{Colour online: Plot of $P_{j0}(t)$ for a $3$-site ring, for
a particle initially on site $1$, in the strong decoherence limit.
Left: The probability to occupy site $0$ (full line), $1$ (large
dashes), and $2$ (small dashes). Right: the current from site $0$ to
site $1$ (compare Fig. \ref{1}). The results do not depend on
$\Phi$.} \label{2}
\end{figure}

As $\lambda$ increases, $p^{}_{max}$ decreases, and Eq. (\ref{Pjaa})
applies at shorter times. Remarkably, if $\lambda> 0.1$ the whole
sum becomes negligible, and we have already reached the strong
decoherence result where the result is $\Phi-$independent. The
result is shown in Fig. \ref{2}. Thus, if we define the 'strong
decoherence' regime as that where all results are flux-independent,
then it is reached for very low values of $\lambda$. We emphasize
here that the detailed form of the results, as well as the
decoherence strength required for flux-independent dynamics, depends
strongly on the form we adopt for either $F_0(p)$ or $\bar{F}(p)$;
we do not have space to explore this question here.

Turning now to the current through the ring, we generalize the free
particle results in the same way as above. Quite generally one has
\begin{equation}
I_{j,j+1}(t)\;=\;
i\langle\tilde{\Delta}_{j,j+1}\rho_{j+1,j}(t)-
\tilde{\Delta}_{j+1,j}\rho_{j,j+1}(t)\rangle \label{54}
\end{equation}
where we average the operator
\begin{equation}
\tilde{\Delta}_{j,j+1}=\Delta_o e^{i \Phi/N} e^{i\sum_k
\mbox{\boldmath $\alpha$}_k^{j,j+1}\cdot\mbox{\boldmath $\sigma$}_k}\label{55}
\end{equation}
over bath states, with fixed bath couplings - the case where one
also averages over an ensemble of bath couplings is a by now obvious
generalization of this. This expression is evaluated in detail in
the Appendix; as noted there, the result is more complicated than it
seems, because the density matrix depends implicitly on both the
initial state, and on the full details of the propagator for the
density matrix. Here we consider only the special case where the
particle starts from the origin, and the fully symmetric case
$\mbox{\boldmath $\alpha$}_k^{ij} \rightarrow \mbox{\boldmath
$\alpha$}_k$. Then one has, for the case of a bath state average
only, that
\begin{widetext}
\begin{align}
I_{j,j+1}(t) \;=\;\frac{2\Delta_o}{N}\sum_{m=0}^{N-1}\sum_p
J_{Np+1}(4\Delta_o t\sin\frac{k_m}{2})
e^{-ik_m(\frac{Np+1}{2}+j)}i^{Np+1}F_0(p)\cos[(\frac{\pi}{2}N+\Phi)p]
 \label{I-iso}
\end{align}
\end{widetext}
with a similar result for the current $\bar{I}_{j,j+1}(t)$ arising
in the case where one also averages over bath couplings, with
$F_0(p)$ then replaced by $\bar{F}_0(p)$.  One can also analyze this
result as a function of time, and of the decoherence strength, the
ring size, and the flux - there is no space for this here. To
nevertheless give some flavour for the results, consider again the
3-site ring, for the coupling averaged case, in the strong
decoherence limit. Current then only flows in regions where the
initial density matrix is inhomogeneous; for some general initial
density matrix one finds
\begin{align}
\bar{I}_{j,j+1}(t) \rightarrow
\frac{2\sqrt{3}}{3}\Delta_o(\rho^{(in)}_{j,j} -
\rho^{(in)}_{j+1,j+1})J^{}_1(2\Delta_o \sqrt{3} t)\ ,
\end{align}
where $\rho^{(in)}_{ll'}$ is the initial density matrix. Again we
see that the result is completely independent of the flux.



\subsection{Comparison with Oscillator Bath} \label{sec:OscB-B}


To gain some perspective on the results just given, it is useful to
compare with what one might expect for a ring particle coupled to an
oscillator bath. The differences are both formal and physical, and
both are important. Here we simply sketch these - a more detailed
study of this rather complex problem will appear elsewhere
\cite{zhen3}. To specify the formal problem completely, one needs
first to define 'spectral functions' for the couplings between the
oscillator bath and the ring particle \cite{cal83}. These couplings
were defined earlier, in (\ref{V-trans}); Fourier transforming them
in the same way as we did for the spin bath couplings, we then
define the spectral functions as:
\begin{align}
J_{\bf p}^{\perp}(\omega) = & {\pi \over 2} \sum_q {U_q^2({\bf p})
\over \omega_q} \delta(\omega - \omega_q)
 \nonumber \\
J_{\bf p}^{\parallel}(\omega) = & {\pi \over 2} \sum_q {V_q^2({\bf
p}) \over \omega_q} \delta(\omega - \omega_q)
 \label{J-kw}
\end{align}
In many cases the non-diagonal function $J_{\bf p}^{\perp}(\omega)$
can be neglected compared to the diagonal $J_{\bf
p}^{\parallel}(\omega)$, and we will assume this here. $J_{\bf
p}^{\parallel}(\omega)$ can take many forms; the most commonly
analysed is the "Ohmic form", where $J_{\bf p}^{\parallel}(\omega) =
\eta \omega$ at low frequency, but this form is very useful for
systems coupled to an itinerant electron bath, it is inappropriate
for insulating systems (where a more accurate low-$\omega$ form is
the "superOhmic" form $J_{\bf p}^{\parallel}(\omega) \sim \omega^k$,
with $k
> 1$). In addition, there is often significant low-energy structure in
$J_{\bf p}^{\parallel}(\omega)$, not describable by a simple
power-law form; and in many cases $J_{\bf p}^{\parallel}(\omega)$
also depends strongly on temperature $T$.

Defining the influence functional ${\cal F}[\Theta, \Theta']$ in the
usual way for general paths $\Theta(t), \Theta'(t)$ (cf. eqtn.
(\ref{infF}) of the Appendix), we can write
\begin{align}
 {\cal F}&[\Theta, \Theta'] = \; \exp {iN^2 \over \hbar} \int dt_1 \int dt_2 \; \times
 \nonumber \\
 &[\dot{\varphi}(t_1) D_{\bf p}(t_1 - t_2) \dot{\varphi}(t_2) + i \Gamma_{\bf p}(t_1
 - t_2) \dot{\varphi}(t_1) \dot{\psi}(t_1)]
 \label{F-oscR}
\end{align}
where we have defined the sum and difference angular variables
\begin{align}
\psi(t) = & \;(\Theta(t) + \Theta'(t))/2
 \nonumber \\
\varphi(t) = & \;(\Theta(t) - \Theta'(t))/2
 \label{psiphi}
 \end{align}
and the oscillator propagator ${\cal D}_{\bf p}(t) = D_{\bf p}(t) +
i \Gamma_{\bf p}(t)$, with
\begin{align}
D_{\bf p}(t) = \;& {4 \over N^2}  \int d \omega \;{J_{\bf
p}^{\parallel}(\omega) \over \omega^2} (1 - \cos \omega t) \coth
({\beta \hbar \omega \over 2})
 \nonumber \\
 \Gamma_{\bf p}(t) = \;& {4 \over N^2}  \int d \omega \;
 {J_{\bf p}^{\parallel}(\omega) \over \omega^2} \sin \omega t
 \label{DGamma}
\end{align}
The behaviour in time of $D_{\bf p}(t)$ can be quite complex, and
varies strongly with the form of $J_{\bf p}(\omega)$, and with
temperature; the details of this behaviour have been reviewed
extensively \cite{weiss99,ajl87}.

In the same way as for the spin bath, we may now construct
expressions for the reduced density matrix, and physical correlation
functions derived therefrom, by summing over all paths; this is done
in a simple generalisation of methods developed for the spin-boson
\cite{ajl87} and Schmid \cite{schmid83,mpaf85} models. For example,
the probability $P_{n0}(t)$ takes the form
\begin{widetext}
\begin{align}
 P_{n0}(t) \;=\; \sum_{p=-\infty}^{\infty} \sum_{l = |n + Np|}^{\infty}
 (-1)^{l-n-Np} e^{i \Phi(p + n/N)} \Delta_o^{2l} \int_0^t dt_{2l}
 \int_0^{t_{2l}} ...\int_0^{t_1} \sum_{\{ q_r \}} \sum_{ \{ \sigma_r
 \}} F(\{ q_r \}, \{ \sigma_r \}; \{ t_l \})
 \label{Pn0-osc}
\end{align}
\end{widetext}
written as a sum over winding numbers $p$ and the number of
intersite hops $l$. In this expression the influence functional has
now become a function $F(\{ \xi_r \}, \{ \chi_r \}; \{ t_l \})$ of
the times $t_l$ at which the particle hops, and of two sets of
'charges' $\{ q_r \} = \pm 1$, $\{ \sigma_r \} = \pm 1$. These
charges are defined in terms of the sum and difference paths by
\begin{align}
\psi(t) = \;& {\pi \over N} \sum_{r=1}^l q_r \theta (t - t_r)
 \nonumber \\
\varphi(t) =\;& {\pi \over N} \sum_{r=1}^l \sigma_r \theta (t - t_r)
 \label{qsigma}
\end{align}
so that the $q_r$ describe hops in the 'centre of mass' part of the
density matrix, and the $\sigma_r$ are hops in the 'difference' or
off-diagonal elements of the density matrix. The general form of
$F(\{ \xi_r \}, \{ \chi_r \}; \{ t_l \})$ is
\begin{align}
F( & \{ \xi_r \}, \{ \chi_r \}; \{ t_l \}) \;=\; \delta(2n -
\sum_{r=1}^{2l} q_r)\; \delta(\sum_{r=1}^{2l} \sigma_r)
 \nonumber \\
&\times \;  \exp {i \over \hbar} \sum_{r' < r}[D_{\bf p}(t_r -
t_{r'})q_r \sigma_{r'} + i \Gamma(t_r - t_{r'}) \sigma_r
\sigma_{r'}]
 \label{F-exp}
\end{align}
and we get the well-known oscillator-mediated interactions between
the charges, familiar from the spin-boson and Kondo problems. Thus
from the formal point of view, for a ring particle coupled to either
an oscillator or spin bath, the principal difference between the two
cases is the existence, in the oscillator bath case, of retarded
interactions between particle hops at different times, whose form
depends on $J_{\bf p}^{\parallel}(\omega)$ and on $T$. Just as in
the spin-boson and Schmid models, the interactions between the
charges in the Ohmic case eventually cause a zero temperature
Kosterlitz-Thouless binding transition between the charges, which
localizes the particle at one site in the ring. This happens at a
critical Ohmic coupling strength $\eta \rightarrow \eta_c = \hbar /2
\pi$, independent of the ring size \cite{zhen3}. When $\eta <
\eta_c$, the particle dynamics is strongly diffusive - we do not go
here into the details of how the dynamics varies with $\eta$, with
$N$, and with temperature $T$. In the superOhmic case there is no
localization transition, no matter how strong the coupling; the
analysis of this case is very lengthy \cite{zhen3}.

None of these features has any formal counterpart in the coupling to
a spin bath. In the present case the spin bath results are entirely
independent of $T$, because all bath levels are degenerate. Even
when this is not the case (ie., when we add back the local fields
$\{ {\bf h}_k \}$, so that the decoherence becomes temperature
dependent), the only way that interactions can be generated between
different bath spins is through their coupling to the particle
itself - there is no analogue to the propagator ${\cal D}_{\bf
p}(t)$.

The key physical difference between this ring-oscillator bath model,
and the ring coupled to a spin bath, is that in the oscillator bath
system, decoherence is in a certain sense a mere side effect of the
dissipation taking place each time the particle excites an
oscillator. On the other hand in the spin bath model, no such
dissipation occurs, only phase decoherence. This difference is most
obviously seen in the centre of mass dynamics of a particle
wave-packet - for the oscillator bath model a wave-packet initially
moving around the ring will dissipate centre of mass momentum
(formally this happens via the interactions between $q_r$ and
$\sigma_{r'}$ in the influence function), slowly bringing it to
rest. However, as we see in the next section, for a ring particle
coupled to a spin bath, the centre of mass momentum of a wave-packet
is completely conserved, even in the strong decoherence limit,
provided the spin bath dynamics is governed by its coupling to the
ring particle (the typical case). This leads to some
counter-intuitive features, as we now see.


\section{ Wave-Packet Interference}
\label{sec:packet}


It is interesting to now turn to the situation where two signals are
launched at $t=0$ from 2 different points in the ring. The idea is
to see how the spin bath affects their mutual interference, and how,
by effectively coupling to the momentum of the particle, it destroys
the coherence between states with different momenta. We do not give
complete results here, but only enough to show how things work.

We therefore start with two-wave-packets which will initially be in
a pure state, and will then gradually be dephased by the bath. In
the absence of a bath, we will assume the wave function of this
state to be the symmetric superposition
\begin{equation}
 \Psi(t) = {1 \over \sqrt{2}}(\psi_1(t) + \psi_2(t))
 \label{Psi1}
\end{equation}
where the two wave-packets are assumed to have Gaussian form:
\begin{eqnarray}
|\psi_1(t)\rangle =& {1 \over Z} &\sum_{n=0}^{N-1}e^{-(k_n-\pi/2)^2
D/2}
\nonumber\\
 \;\;\;\;\;\;\;\;\;\;\;& \times &
 e^{-i j_0 k_n-i 2\Delta_0 t\cos(k_n-\Phi/N)}|k_n\rangle
 \label{psi1}
\end{eqnarray}
\begin{eqnarray}
|\psi_2(t)\rangle=&\frac{1}{Z}&\sum_{n=0}^{N-1}e^{-(k_n-\pi/2)^2
D/2}\nonumber\\
 \;\;\;\;\;\;\;\;\;\;\;& \times & e^{-i 2\Delta_0 t\cos(k_n-\Phi/N)}|2\pi-k_n\rangle
 \label{psi2}
\end{eqnarray}
where we assume the usual symmetric ring with flux $\Phi$, and
$Z=\sqrt{\sum_{n=0}^{N-1}e^{-(k_n-\pi/2)^2 D}}$ is the wave-function
normalization factor. At $t=0$, one of the packets is centred at the
origin, and the other at site $j_o$, and they both have width $D$.
Note that the velocity of each wave-packet is conserved, and at
times such that $\Delta_o t= 2 n$, they cross each other. From
(\ref{psi1}) we see that the main effect of the flux is to shift the
relative momentum of the wave-packets. It also affects the rate at
which the wave-packets disperse in real space - this dispersion rate
is at a minimum when $\Phi/N=\frac{\pi}{2}$.

The free-particle wave function in real space is then
\begin{equation}
\begin{split}
|\Psi_j(t) \rangle=&
\frac{1}{Z\sqrt{2N}}\sum_{n=0}^{N-1}e^{-(k_n-\pi/2)^2D/2}\\&\times(e^{i
(j-j_0) k_n}e^{-2 i \Delta_o t \cos{(k_n+\Phi/N)}}
\\&\;\;\;\;+e^{- i j k_n}e^{-2 i \Delta_o t \cos{(k_n-\Phi/N)}})|j\rangle
\end{split}
\end{equation}
so that the probability to find a particle at time $t$ on site $j$
is $P(j) = |\Psi_j(t)|^2$.

Let us now consider the effect of phase decoherence from the spin
bath. Using the results for $P_{jj'}(t)$ from the last section, with
an initial reduced density matrix
\begin{equation}
\rho^{(in)}_{jj'} \;=\; |\Psi_j(t=0) \rangle \langle \Psi_{j'}(t=0)
|
 \label{rho-0}
\end{equation}
we find a rather lengthy result for the probability that the site
$j$ is occupied at time $t$:
\begin{widetext}
\begin{equation}
\begin{split}
P_j(t)={1 \over 2NZ^2} \sum_{n,n'=0}^{N-1}\sum_{m=-\infty}^{+\infty}
&e^{-((k_n-\pi/2)^2+(k_{n'}-\pi/2)^2) D/2}F_0(m)\\
\times&\{e^{i(j-j_0)(k_n-k_{n'})}J_{m}(4 \Delta_o t
\sin{((k_n-k_{n'})/2)})e^{im((k_n+k_{n'})/2+\Phi/N)}+\\
&+e^{-i (k_n-k_{n'})j}J_{m}(4 \Delta_o t \sin{((k_n-k_{n'})/2)})
e^{im((k_n+k_{n'})/2-\Phi/N)}+\\
&+[e^{i ((j-j_0) k_n+j k_{n'})}J_m(4 \Delta_o t
\sin{((k_n+k_{n'})/2)})e^{im((k_n-k_{n'})-\Phi/N)}+H.c.]\}
\end{split}
 \label{Pj0wp}
\end{equation}

One can also, in the same way, derive results for the current in the
situation where we start with 2 wave-packets. We see that
expressions like (\ref{Pj0wp}) are too unwieldy for simple analysis.
However in the strong decoherence limit (\ref{Pj0wp}) simplifies to:
\begin{align}
 P_j(t)={1 \over 2NZ^2}&\sum_{n,n'=0}^{N-1}e^{-((k_n-\pi/2)^2+(k_{n'}-\pi/2)^2)D/2}
\{e^{i(j-j_0)(k_n-k_{n'})}J_0(4 \Delta t \sin{((k_n-k_{n'})/2)})+
\nonumber\\ &+e^{-i j(k_n-k_{n'})}J_0(4 \Delta t
\sin{((k_n-k_{n'})/2)})+[e^{i ((j-j_0) k_n+j k_{n'})}J_0(4 \Delta t
\sin{((k_n+k_{n'})/2)})+H.c.]\}
\end{align}
\end{widetext}
and again we see that the flux has disappeared from this equation.

\begin{figure}
\includegraphics[width=8cm]{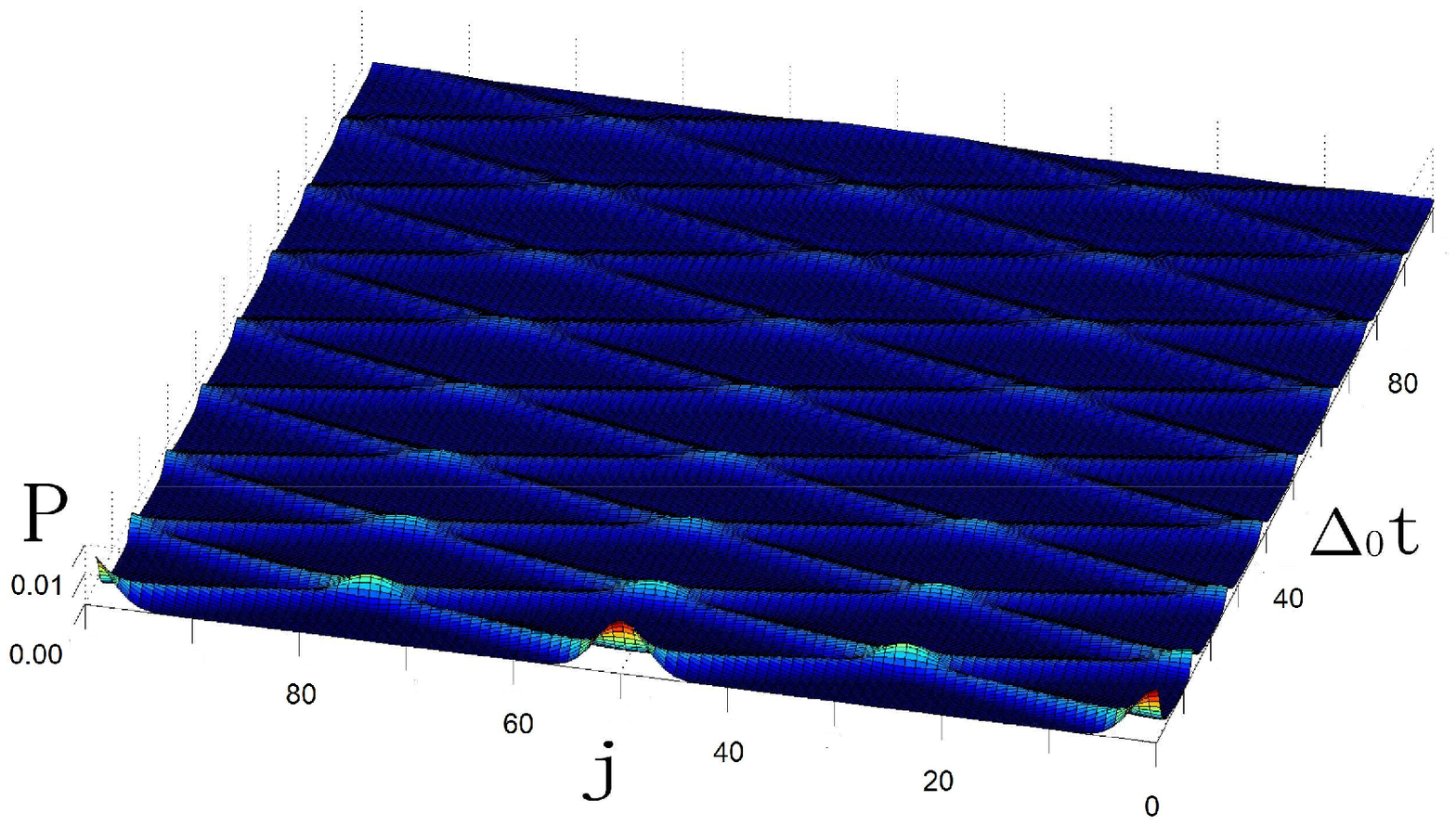}
\includegraphics[width=8cm]{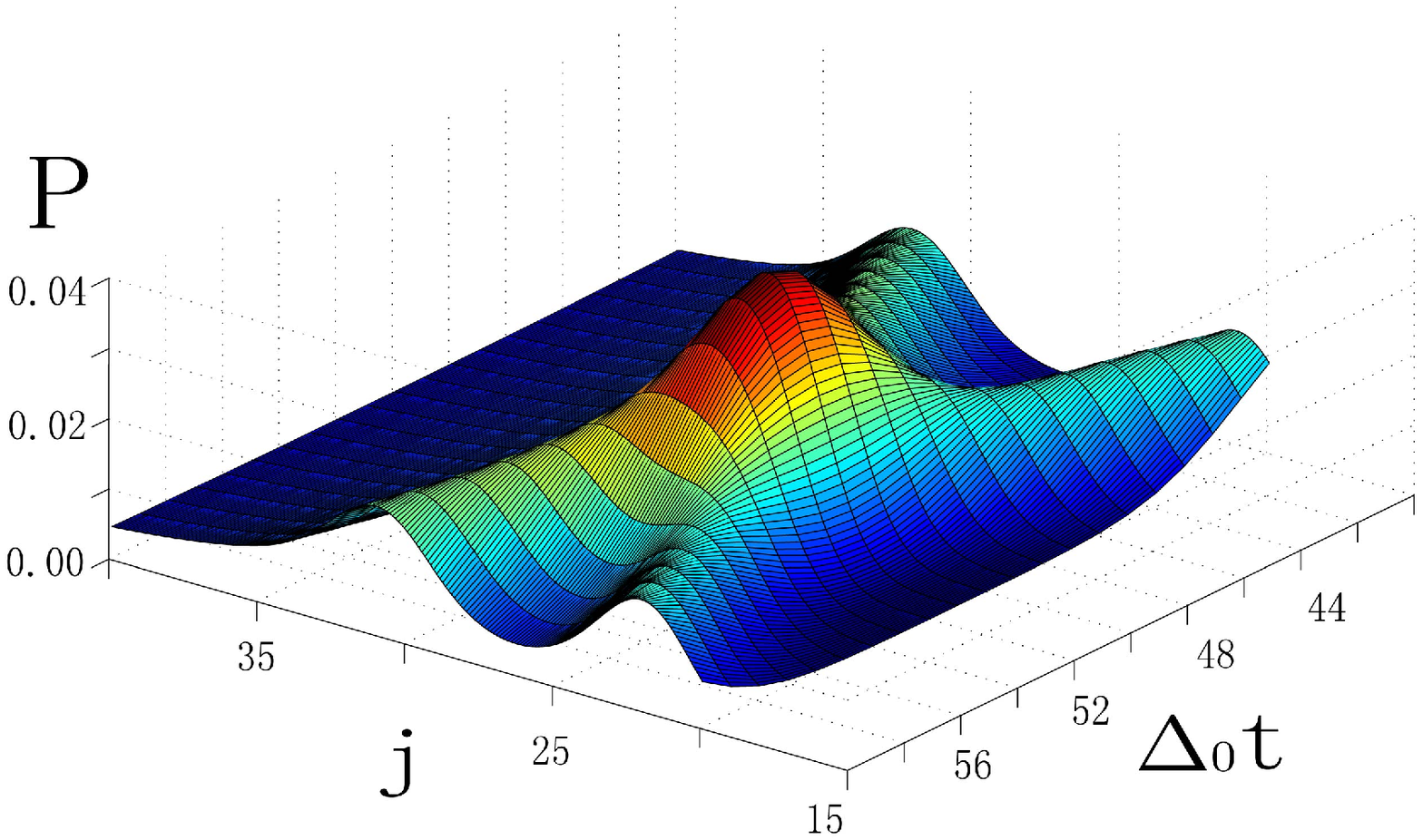}
\caption{\label{fig:wavepacket_with_topological} Colour online: Plot
for $P_j(t)$ as a function of both $j$ and $\Delta_o t$ in the
strong decoherence limit. $j_o = 50$ and $N=100$.  The relative
velocity is $\frac{\pi}{2}$, in phase units. Top: global view.
Bottom: a particular peak}
\end{figure}

This result is shown in Fig. \ref{fig:wavepacket_with_topological}.
As one might expect, the interference between the two wave-packets
is completely washed out in this strong decoherence limit. However
there is also a more unexpected feature - each wave-packet now has
portions moving in opposite directions to each other. The
explanation is to be found by noticing that as the particle hops, at
the same time causing the bath spins to make transitions, the
topological phase it exchanges with the spins also changes the total
phase around the ring seen by the particle. Thus, from the point of
view of the particle, these transitions are forcing the total flux
through the ring to fluctuate, in a way which depends on the
trajectory followed by the particle. This dependence, such that the
changing phase is conditional on the particle path, is of course why
we get decoherence. Now the changing effective flux changes the
particle momentum and velocity, and in the case of the pair of
wave-packets here, it also changes their relative momentum. Indeed,
given that the transformation $\Phi\rightarrow\Phi+\pi$ completely
reverses the momentum, we see that a strong coupling to the bath
spins can even cause a part of the initial wave-packets to reverse
its direction. However we emphasize that the centre of mass momentum
for the combined wave-packet system has {\it not} changed - the
average momentum imparted to the ring particle is zero. Thus, as
noted in the last section, the decoherence caused by the bath is not
accompanied by any net dissipation of the particle momentum, or of
its energy. Indeed, if a wave-packet starts off with a net angular
momentum around the ring, this will be conserved, long after all
coherence has been lost.

Note that these results are {\it not} the same as one would get by
just adding a fluctuating noise $\delta \Phi(t)$ to the static flux.
Such an external noise term will also cause 'noise' decoherence, but
of a quite different form from that of the spin bath decoherence
discussed here, since there is no correlation between the noise and
the particle dynamics. In fact, as we will discuss elsewhere, a
fluctuating flux noise acting on the particle causes exponential
decay in time of the particle correlation functions, quite different
from the power law decay typical for the present case.


\section{Summary \& Conclusions}
 \label{sec:summary}


Let us first recall the main results derived in sections II-V. In
section II we show how the basic Hamiltonian (\ref{eq:H_phi'}) we
have studied can be derived, including the fairly severe
approximations that are involved. No attempt is made to connect the
model with any specific physical system, since the main focus of
this paper is to study pure phase decoherence in a solvable model.
The most important features of the model are that (i) the spin bath
which couples to the model can cause severe phase decoherence with
no dissipation, and (ii) the phase interference in the ring
(including Aharonov-Bohm oscillations) is affected in a rather
fascinating way by the decoherence; and (iii) the model can be
solved exactly. The main tasks we set ourselves in this paper were
to set up a formal apparatus to solve this model, and to study some
aspects of the decoherence dynamics in it.

Before studying the decoherence, it turns out to be important to
develop the detailed solution for the dynamics of the $N$-site ring
without the bath, in section III - to our surprise, this does not
seem to have been done before. It is convenient to develop the free
particle density matrix $\hat{\rho}^o(t)$ and its propagator
$\hat{K}^o(t)$ as double sums over winding numbers around the ring
(see eqtn. (\ref{rho-oA}) for the propagator); but we also show how
this can be rewritten as a single sum over winding numbers (see
eqtn. (\ref{rho1B}), a form more useful for numerical work on large
rings.

The importance of the work on the free particle problem is seen in
the result (\ref{K-F}) for the propagator $\hat{\cal K}(t)$ of the
reduced density matrix once one integrates out the spin bath - we
can find this by summing over winding numbers an expression
involving matrix elements of $\hat{K}^o(t)$ and a weighting function
$F_{jj'}^{ll'}(p,p')$, the influence function. The influence
function can be found exactly (Appendix A.2); we do this both for
the case where the couplings between the particle and the bath spins
are fixed, and the case where we make an ensemble average over these
couplings. This allows us to derive a whole series of exact
expressions for the propagator $\hat{\cal K}(t)$ (see Appendix A.2,
eqtns. (\ref{K-av1}) and (\ref{K-av2})), and thence for the time
evolution of the reduced density matrix, the probability density,
and the current (section IV). In deriving results for these physical
quantities, one finds that the full details of $F_{jj'}^{ll'}(p,p')$
are not required, but only certain matrix elements (often only the
element $F_0(\bar{p})$, produced by letting $\mu = j-j'+l-l' = 0$,
and $p'-p = \bar{p}$; see (\ref{F_0'}) {\it et seq.}). The
characteristics of the decoherence are controlled by these.

It turns out that the decoherence dynamics, and how it affects
different physical quantities, depends on the ring size $N$, the
flux $\Phi$, and the form of the function $F_0(\bar{p})$; and also
on the initial state of the system. A full exploration of this large
parameter space would take a lot of space, so we have focussed on
certain questions. One of these is the flux dependence of the
various physical quantities, and how phase decoherence affects these
- how this works is shown in section IV, with details given for a
3-site ring. We also look at the interference of 2 wave-packets on a
large ring, in section V. This of course depends crucially on the
flux, and as we switch on decoherence, this flux dependence
disappears, even though the wave-packets still propagate (although
the coupling to the spin bath also strongly distorts the shape of
the wave-packets). In all cases we find that the detailed dynamics
is quite different from what one would get if the decoherence was
simulated by adding flux noise to the problem - in particular, all
coherence properties show power law decay in time, instead of
exponential decay.

Because of the size of the parameter space, there is much that is
unexplored in this paper - in particular, we expect the decoherence
dynamics, and its dependence on flux, to depend very dramatically on
the form of $F_0(\bar{p})$; and we have hardly explored the
dependence on ring size. Nor have we attempted any connection to
experiment. The main reason for this is that for a detailed
comparison with experiments on most real systems one has to add two
crucial ingredients, viz. (i) we must add back the local fields $\{
{\bf h}_k \}$ acting on the bath spins, and the diagonal couplings
$\{ \mbox{\boldmath $\gamma$}_{k}^{j} \}$ (compare eqtns.
(\ref{eq:H_SB}) and (\ref{H-band1})); and (ii) in many physical
applications there are also important couplings to delocalised modes
like phonons, which are modeled using oscillator bath interactions.
Actually one can also solve the problem when these couplings are
added, in certain parameter ranges - this will be the subject of
future papers \cite{zhen3}.


\begin{acknowledgments}

AA, OE-W, and PCES acknowledge the support of PITP for this work. ZZ
and PCES were also supported by NSERC, and PCES by CIFAR, in Canada,
and AA and OE-W by the German-Israeli project cooperation (DIP) and
the US-Israel binational foundation.

\end{acknowledgments}


\appendix


\section{}
 \label{sec:App}


In this Appendix we derive some of the expressions for Green
functions and density matrices that are used in the text, and also
explain some of the mathematical transformations required to go from
single sums over winding number to double sums.


\subsection{ Free Particle}
 \label{sec:App-A1}


We consider first the free particle for the $N$-site symmetric ring,
with Hamiltonian
 \begin{equation}
 \label{eq:H-o}
H_o \;=\;\sum_{<ij>} \left[ \Delta_o c_i^{\dagger} c_j \: e^{i
\Phi/N} + H.c.\right]
 \end{equation}
and band dispersion $\epsilon_{k_n} = 2\Delta_o \cos(k_n - \Phi/N)$.

For this free particle the dynamics is entirely described in terms
of the bare 1-particle Green function
\begin{align}
 G^o_{jj'}(t) &\equiv \langle j|G^o(t)|j'\rangle
\equiv\langle j|e^{-i{\cal H}_ot}|j'\rangle\nonumber\\
&=\frac{1}{N}\sum_{n}e^{-i2\Delta^{}_{0}t\cos (k^{}_{n}-\Phi/N
)}e^{ik^{}_{n}(j' -j)}\ .
 \label{GLM2}
\end{align}
which gives the amplitude for the particle to propagate from site
$j'$ at time zero to site $j$ at time $t$. This can be written as a
sum over winding numbers $m$, viz.,
\begin{align}
G^{o}_{jj'}(t) =\sum_{\ell =0}^{\infty}&\sum_{m=0}^{\ell}
\frac{(-i\Delta^{}_{o}t)^{\ell}}{m !(\ell -m)!}e^{i\Phi/N (\ell -2m)}\nonumber\\
&\times\frac{1}{N}\sum_{n=0}^{N-1}e^{-i\frac{2\pi n (\ell
-2m-j+j')}{N}}\
\end{align}
This sum may be evaluated in various forms, the most useful being in
terms of Bessel functions:
\begin{equation}
\begin{split}
&G^o_{jj'}(t)\\
 &=\frac{1}{N}\sum_{n=0}^{N-1} \sum_{m=-\infty}^{+\infty}
J_{m}(2\Delta_o t) (-i)^{m} e^{i m (k_n-\Phi/N)+i k_n(j-j')}\\
 &=\sum_{m=-\infty}^{+\infty}J_{m}(2\Delta_o t) (-i)^{m}
e^{-i m \Phi/N}\delta_{Np, m+j-j'}\\
 &=\sum_{p=-\infty}^{+\infty}J_{Np+j'-j}(2\Delta_o t) e^{-i (Np+j'-j)
(\Phi/N+\pi/2)}
\end{split}
 \label{Bess-2sum}
\end{equation}
(this last form, where we have eliminated the sum over winding
numbers, is also of course directly derivable from (\ref{GLM2})). We
can also write this last form as
\begin{equation}
G^{o}_{jj'}(t) = \sum_{p}e^{ip\Phi +i{\Phi \over N}
(j-j')}I_{Np+j-j'}(-2i\Delta^{}_{o}t)\ ,
 \label{GJJ1}
\end{equation}
where we use the hyperbolic Bessel function $I_{\alpha}(x)$, defined
as $I^{}_{\alpha}(x)=(i)^{-\alpha}_{}J^{}_{\alpha}(ix)$.

Consider now the free particle density matrix. As discussed in the
main text, we have in general some initial density matrix
$\rho^{(in)}_{l,l'} = \langle l |\hat{\rho}^o(t=0)|l' \rangle$ at
time $t=0$ (where $l$ and $l'$ are site indices). Then at a later
time $t$ we have
\begin{equation}
\rho^o_{jj'}(t) = \sum_{l,l'} K^o_{jj',ll'}(t) \rho^{(in)}_{l,l'}
\label{rho-ot2}
\end{equation}
where $K^o_{jj',ll'}(t)$ is the propagator for the free particle
density matrix. Its form follows directly from the definition of
this density matrix as $\hat{\rho}^o(t) = |\psi(t)\rangle \langle
\psi(t)|$, where $|\psi(t) \rangle$ is the particle state vector at
time $t$. One thus has
\begin{equation}
K^o_{jj',ll'}(t) =  G^o_{jl}(t)
G^o_{j'l'}(t)^{\dagger}.\label{rho-ot2b}
\end{equation}

An obvious way of writing this propagator is then:
\begin{equation}
K^o_{jj',ll'}(t) = \sum_{pp'} K^o_{jj',ll'}(p,p';t)
 \label{Ko-Kopp'}
\end{equation}
where we have a double sum over winding numbers $p,p'$. The explicit
form for $K^o_{jj',ll'}(p,p';t)$ is then given from
(\ref{Bess-2sum}) as
\begin{widetext}
\begin{align}
K^o_{jj',ll'}(t)&=\sum_{pp'}e^{i(p-p')\Phi}e^{i\Phi (j-j'+
l-l')/N}i^{-Np -j + l}i^{Np'+j'-l'}
J^{}_{Np+j-l}(2\Delta^{}_{o}t)J^{}_{Np'+j'-l'}(2\Delta^{}_{o}t) \nonumber\\
&=\sum_{pp'}e^{i(p-p')\Phi}e^{i\Phi (j-j'+ l-l')/N}
 I^{}_{Np+j-l}(-2i\Delta^{}_{o}t)I^{}_{Np'+j'-l'}(2i\Delta^{}_{o}t)\
 .
  \label{rho-oA}
\end{align}
However this expression is somewhat unwieldy, particularly for
numerical evaluation, because of the sum over pairs of Bessel
functions. It is then useful to notice that we can also derive the
answer as a single sum over winding numbers, as follows:
\begin{align}
&K^o_{jj',ll'}(t)=\frac{1}{N^2}\sum_{n,n'=0}^{N-1}
e^{-i(k^{}_n(j-l)-k^{}_{n'}(j'-l')) +4 i
\Delta^{}_ot\sin[\Phi/N-(k^{}_n+k^{}_{n'})/2]\sin[(k^{}_{n}-k^{}_{n'})/2]}
\nonumber\\
&=\frac{1}{N^2}\sum_{n,m=0}^{N-1}\sum_{p=-\infty}^\infty
J^{}_p[4\Delta^{}_ot\sin(k^{}_m/2)] e^{i
p(\Phi/N-k^{}_{n}+k^{}_m/2)-ik^{}_n(j-l)+i(k^{}_{n}-k^{}_m)(j'-l')}\nonumber\\
&=\frac{1}{N} \sum_{p'=-\infty}^\infty \sum_{m=0}^{N-1}\bigl
(J^{}_{Np'+j'-j+l-l'}[4\Delta^{}_o t \sin
(k^{}_m/2)]e^{ik^{}_m(l+l'-j-j'+Np')/2}\bigr )e^{i\Phi/N
(Np'+j'-j+l-l')}\ .\label{rho1B}
\end{align}
\end{widetext}
In the second step we replaced $n'=m-n$. In the third step we also
used the identity $\sum_{n'=0}^{N-1}e^{i k^{}_{n'}\ell}\equiv
\sum_{p'=-\infty}^\infty N\delta_{\ell,Np'}$.

The result for the density matrix then depends on what is the
initial density matrix, according to (\ref{rho-ot2}). If we start
with $\hat{\rho}^{(in)} = |0\rangle\langle 0|$, the density matrix
is then just $\rho^o_{jj'}(t) = K^o_{jj',00}(t)$. One then gets a
much simpler expression; the density matrix at time $t$ is:
\begin{equation}
 \begin{split}
\rho^o_{jj'}(t) =\frac{1}{N}&\sum_{m=0}^{N-1}
\sum_{p'=-\infty}^\infty J^{}_{Np'+j'-j}[4\Delta^{}_o t \sin
(k^{}_m/2)] \\
&\times e^{i\phi(Np'+j'-j)-ik^{}_m(j+j'-Np')/2}
 \label{rho-00}
 \end{split}
 \end{equation}

It is useful and important to show that the double- and single-sum
expressions (\ref{rho-oA}) and (\ref{rho1B}) are equivalent to each
other. To do this we use Graf's summation theorem for Bessel
functions \cite{Graf}, in the form:
\begin{equation}
J_{\nu}(2x\sin\frac{\theta}{2})(-e^{-i\theta})^{\frac{\nu}{2}}=
\sum_{\mu=-\infty}^{+\infty}J_{\nu+\mu}(x)J_{\mu}(x)e^{i\mu\theta}
\end{equation}
We set $\theta=0,\frac{2\pi}{N},...\frac{2\pi
m}{N},...\frac{2\pi(N-1)}{N}$, which is the $k_m$ in \eqref{rho1B}
and multiply by $e^{-i\theta j}$ on each side. We then have
\begin{align}
J_{\nu}&(2x\sin\frac{k_m}{2})e^{-i(k_m+\pi)\frac{\nu}{2}}e^{-i k_m j} \nonumber\\
&=\sum_{\mu=-\infty}^{+\infty}J_{\nu+\mu}(x)J_{\mu}(x)e^{i(\mu-j)k_m}
 \end{align}

Noticing then that
$\sum_{m=0}^{N-1}e^{i k_m n} = N\sum_p \delta_{Np, n}$
we then do the sum over $m$; only $\mu-j=Np$ survives, and thus
\begin{align}
\frac{1}{N}&\sum_{m=0}^{N-1}J_{\nu}(2x\sin\frac{k_m}{2})
e^{-i(k_m+\pi)\frac{\nu}{2}}e^{-i
k_m j} \nonumber\\
 &=\frac{1}{N}\sum_p J_{Np+j+\nu}(x)J_{Np+j}(x)
\end{align}
Setting $\nu=Np'+j'-Np-j + l - l'$, $x=2\Delta_o t$, we then
substitute back into \eqref{rho-oA}, to get
\begin{widetext}
\begin{align}
 \label{rho-o'}
K^o_{jj',ll'}(t)&=\sum_{pp'}e^{i(\Phi/N+\pi/2)(Np'-Np
+j'-j+l-l')}J_{Np+j-l}
(2\Delta^{}_{o}t)J^{}_{Np'+j'-l'}(2\Delta^{}_{o}t) \nonumber\\
&=\frac{1}{N}\sum_p
e^{+i(Np+j'-j+l-l')(\frac{\Phi}{N}+\frac{\pi}{2})}
\sum_{m=0}^{N-1}J_{Np+j'-j+l-l'}
(4\Delta_o t\sin\frac{k_m}{2})e^{-i(k_m+\pi)\frac{Np+j'-j+l-l'}{2}}
e^{-i k_m j} \nonumber\\
&=\frac{1}{N}\sum_p\sum_{m=0}^{N-1}J_{Np+j'-j+l-l'}(4\Delta_o
t\sin\frac{k_m}{2})e^{i(Np+j'-j)\frac{\Phi}{N}-ik_m(j+j'-l-l'+Np)/2}
\end{align}
The propagator $\rho$ is Hermitian, ie., $K^o_{jj',ll'}(t)=
K^o_{j'j,l'l}(t)^{*}$; setting $p'=-p$, we then have
\begin{align}
K^o_{jj',ll'}(t)&=\frac{1}{N}\sum_{p'}
\sum_{m=0}^{N-1}J_{-Np'+j-j'+l'-l}(4\Delta_o t\sin\frac{k_m}{2})
e^{i(Np'+j'-j+l-l')\frac{\Phi}{N}+ik_m(j+j'-l-l'-Np')/2} \nonumber\\
&=\frac{1}{N}\sum_{m=0}^{N-1}\sum_{p'=-\infty}^\infty
J^{}_{Np'+j'-j+l-l'}[4\Delta^{}_o t
\sin(k^{}_m/2)]e^{i\phi(Np'+j'-j+l-l')-ik^{}_m(j+j'-l-l'-Np')/2}
\end{align}
where in the last line, we set $k_m\rightarrow -k_m$, and use the
fact that for integer order $n$, $J_n(-x)=J_{-n}(x)$. Thus we have
demonstrated the equivalence of the single and double sum forms for
the density matrix.
\end{widetext}


\subsection{ Including Phase Decoherence}
 \label{sec:App-A2}


To calculate the reduced density matrix for the particle in the
presence of the spin bath, we need to average over the spin bath
degrees of freedom. We will do this in a path integral technique,
adapting the usual Feynman-Vernon \cite{feynman63} theory for
oscillator baths to a spin bath; the following is a generalization
of the method discussed previously \cite{PS00}. We can parametrize a
path for the angular coordinate $\Theta(t)$ which includes $m$
transitions between sites in the form
\begin{equation}
\Theta_{(m)}(\{ q_i \},t)\;\;=\;\; \Theta(t=0) \;+\; \sum_{i=1}^{m}
q_i \theta(t-t_{i}) \;
 \label{Theta_m}
\end{equation}
where $\theta(x)$ is the step-function; we have transitions either
clockwise (with $q_j = +1$) or anticlockwise (with $q_j = -1$) at
times $t_1,t_2, \dots ,t_{m}$. The propagator ${\cal K}(1,2)$ for
the particle reduced density matrix between times $\tau_1$ and
$\tau_2$ is then
\begin{equation}
{\cal K}(1,2) \ = \displaystyle \int^{\Theta_2}_{\Theta_1} d\Theta
\displaystyle \int^{\Theta^{\prime}_{2}}_{\Theta^{\prime}_{1}}
d\Theta^{\prime} \ e^{-{i \over \hbar} (S_o[\Theta]  -
S_o[\Theta^{\prime}])} {\cal F}[\Theta,\Theta'] \;
 \label{K12}
\end{equation}
where $S_o[\Theta]$ is the free particle action, and ${\cal
F}[\Theta, \Theta^{\prime}]$ is the ``influence functional"
\cite{feynman63}, defined by
\begin{equation}
{\cal F}[\Theta,\Theta'] =  \prod_{k} \langle \hat{U}_k(\Theta,t)
\hat{U}_k^{\dag}(\Theta',t) \rangle \;
 \label{infF}
\end{equation}
Here the unitary operator $\hat{U}_k(\Theta,t)$ describes the
evolution of the $k$-th environmental mode, given that the central
system follows the path $\Theta(t)$ on its "outward" voyage, and
$\Theta'(t)$ on its "return" voyage. Thus ${\cal F}[\Theta,\Theta']$
acts as a weighting function, over different possible paths
$(\Theta(t),\Theta'(t))$. The average $\langle ...\rangle$ is
performed over environmental modes - its form depends on what
constraints we apply to the initial full density matrix. In what
follows we will assume an initial product state for the full
particle/environment density matrix.

For the general Hamiltonian in eqtns. (\ref{eq:H})-(\ref{eq:H_SB}),
the environmental average is a generalisation of the form that
appears \cite{PS00,PS06} when we average over a spin bath for a
central 2-level system, or "qubit". The essential result is that we
can calculate the reduced density matrix for a central system by
performing a set of averages over the bare density matrix. For a
spin bath these can be reduced to phase averages and energy
averages; and for the present case it reduces to a simple phase
average.

To see this, notice that the sole effect of the pure phase coupling
to the spin bath is simply to accumulate an additional phase in the
path integral each time the particle hops. Just as for the free
particle, we can then classify the paths by their winding number;
for a path with winding number $p$ which starts at site $l$ (the
initial state) and ends at site $j$, the additional phase factor can
then be written as
\begin{equation}
 \label{additonal_phase}
\exp\{-i p\sum_k   \left(  \sum_{\langle mn\rangle=\langle
01\rangle}^{\langle N 0\rangle}  - \sum_{\langle mn\rangle=\langle
l,l+1\rangle}^{\langle j-1,j\rangle}\right) (\mbox{\boldmath
$\alpha$}_k^{mn}\cdot\mbox{\boldmath $\sigma$}_k) \}
\end{equation}
and for fixed initial and final sites, this additional phase only
depends on the winding number.

Consider now the form this implies for the reduced density matrix of
the particle, once the bath has been averaged out. The equation of
motion for the reduced density matrix will be written as
\begin{equation}
\rho_{jj'}(t) = \sum_{l,l'} {\cal K}_{jj',ll'}(t) \rho^{(in)}_{l,l'}
\label{rho-ot2'}
\end{equation}
where ${\cal K}_{jj',ll'}(t)$ is the propagator for the reduced
density matrix. Now the key result is that
\begin{equation}
{\cal K}_{jj',ll'}(t) =  \sum_{pp'} K^o_{jj',ll'}(p,p';t)
F_{jj'}^{ll'}(p,p')
 \label{K-F'}
\end{equation}
where the function $K^o_{jj',ll'}(p,p';t)$ is the free particle
propagator for fixed winding numbers $p,p'$ (see eqtn.
(\ref{Ko-Kopp'}) above). This form follows from the argument just
given, viz., that the only effect of the spin bath is to add the
extra phase factor (\ref{additonal_phase}) in each path in the path
integral for the propagator. Thus the influence functional,
initially over the entire pair of paths for the reduced density
matrix, has now reduced to the much simpler weighting function
$F_{jj'}^{ll'}(p,p')$, which we will henceforth call the "influence
function". To evaluate this influence function, we must specify what
kind of bath average we wish to take. We consider here the two cases
discussed in the text, viz., (i) a simple average $<...>$ over bath
states, and (ii) an average $<<....>>$ over both the bath states and
over a distribution $P(\mbox{\boldmath $\alpha$}_k^{mn})$ of
couplings to the bath spins. The results are obtained as follows:

\begin{widetext}
(i) In the case of fixed bath couplings $\mbox{\boldmath
$\alpha$}_k^{mn}$, the average is obtained by simply inserting the
phase factors (\ref{additonal_phase}) from above into the paths of
different winding number.  We then get:
\begin{align}
F_{jj'}^{ll'}(p,p') \; = \; \langle e^{-i (p-p')
 \sum_k\sum_{\langle mn\rangle=\langle
0, 1\rangle}^{\langle N-1, N\rangle} \mbox{\boldmath
$\alpha$}_k^{mn}\cdot\mbox{\boldmath $\sigma$}_k }e^{-i
 \sum_k\sum_{\langle mn\rangle=\langle
l', l'+1\rangle}^{\langle l-1, l\rangle} \mbox{\boldmath
$\alpha$}_k^{mn}\cdot\mbox{\boldmath $\sigma$}_k } \; e^{-i
\sum_k\sum_{\langle mn\rangle=\langle j',j'+1\rangle}^{\langle
j-1,j\rangle} \mbox{\boldmath $\alpha$}_k^{mn}\cdot\mbox{\boldmath
$\sigma$}_k}\rangle
 \label{F-gen}
\end{align}
In the symmetric coupling case where $\mbox{\boldmath
$\alpha$}_k^{mn} \rightarrow \mbox{\boldmath $\alpha$}_k$, this
expression reduces to a much simpler result:
\begin{equation}
 \label{eq:F_avg'}
F_{jj'}^{ll'}(p,p')\;=\;\langle e^{i (\mu + N\bar{p})\sum_k
\mbox{\boldmath $\alpha$}_k\cdot\mbox{\boldmath $\sigma$}_k} \rangle
\end{equation}
where we define
\begin{align}
\mu = j'-j +l'-l \; , \;\;\;\;\;\;\;\;\;\;\;\; \bar{p} = p'-p.
 \label{mu-barp}
\end{align}
If the particle is launched from the origin, this gives the even
simpler result (\ref{eq:F_avg}) quoted in the main text.

We can now give the explicit result for the propagator of the
reduced density matrix in double sum form as
\begin{align}
{\cal K}_{jj'}^{ll'}(t) \;=\;  \sum_{pp'} \langle e^{i (\mu +
N\bar{p})\sum_k \mbox{\boldmath $\alpha$}_k\cdot\mbox{\boldmath
$\sigma$}_k} \rangle \; e^{-i\bar{p})\Phi}e^{-i\Phi
\mu/N}i^{N\bar{p} + \mu}
J^{}_{Np+j-l}(2\Delta^{}_{o}t)J^{}_{Np'+j'-l'}(2\Delta^{}_{o}t)
 \label{K-av1}
\end{align}
Typically the average $<...>$ here will be over a set of thermally
weighted states, but this is not required - in principle one could
average with a non-thermal state of the bath (or even a definite
bath state, eg., one which had been polarized beforehand - in this
case no bath averaging is required at all). If we do assume a
thermal state, then since all bath states are degenerate, and hence
equally populated at any finite $T$, eqtn. (\ref{eq:F_avg'}) reduces
to
\begin{align}
F_{jj'}^{ll'}(p,p')\;=\;   \langle e^{i (\mu + N\bar{p})\sum_k
\mbox{\boldmath $\alpha$}_k\cdot\mbox{\boldmath $\sigma$}_k} \rangle
\;\; \rightarrow \;\;\prod_k \cos((N\bar{p} + \mu) | \mbox{\boldmath
$\alpha$} |)
 \label{F-av1}
\end{align}
These results can also be written in single sum form - we do not go
through the details here. Both forms are fairly easily summed
numerically, even for rather large rings.

\vspace{2mm}

(ii) In the case where we must also average over the distribution of
bath couplings, the same phase factors appear, but now we must
average over the bath spin couplings; this gives instead
\begin{align}
\bar{F}_{jj'}^{ll'}(p,p') \; = \; \prod_k \int & d \mbox{\boldmath
 $\alpha$}_k^{mn}  P(\mbox{\boldmath $\alpha$}_k^{mn}) \nonumber\\
 & \times \langle e^{-i (p-p')
 \sum_{\langle mn\rangle=\langle 0, 1\rangle}^{\langle N-1, N\rangle}
 \mbox{\boldmath $\alpha$}_k^{mn}\cdot\mbox{\boldmath $\sigma$}_k}
 e^{-i \sum_{\langle mn\rangle=\langle l', l'+1\rangle}^{\langle
 l-1, l\rangle} \mbox{\boldmath $\alpha$}_k^{mn}\cdot\mbox{\boldmath $\sigma$}_k }
 \; e^{-i \sum_{\langle mn\rangle=\langle
 j',j'+1\rangle}^{\langle j-1,j\rangle} \mbox{\boldmath $\alpha$}_k^{mn}
 \cdot\mbox{\boldmath $\sigma$}_k}\rangle
 \label{F-gen'}
\end{align}
where we put a bar over the influence function to signify an extra
average over bath couplings; again $<...>$ signifies the average
over bath states, and $P(\mbox{\boldmath $\alpha$}_k^{mn})$ is the
probability weighting for the different bath couplings. Typically it
makes little sense, in this ensemble average, to have any dependence
of the $P(\mbox{\boldmath $\alpha$}_k^{mn})$ on the link ${mn}$, so
that this reduces to
\begin{align}
 \label{F-gen''}
\bar{F}_{jj'}^{ll'}(p,p') \; \equiv \;   \prod_k\langle \langle e^{i (\mu +
N\bar{p}) \mbox{\boldmath $\alpha$}_k\cdot\mbox{\boldmath $\sigma$}_k}
\rangle \rangle   \; = \;  \prod_k\int d \mbox{\boldmath $\alpha$}_k
P(\mbox{\boldmath $\alpha$}_k) \;\langle e^{i (\mu + N\bar{p})
\mbox{\boldmath $\alpha$}_k\cdot\mbox{\boldmath $\sigma$}_k} \rangle
\end{align}
where $\mu = j'-j +l'-l$, and $\bar{p} = p'-p$ as above, and where
we use the fact that the distribution $P(\mbox{\boldmath
$\alpha$}_k)$, defined for the individual bath couplings, reduces in
an ensemble average to a simple average over some coupling strength
$\mbox{\boldmath $\alpha$}$, with weighting $P(\mbox{\boldmath
$\alpha$})$, acting on some representative spin \mbox{\boldmath
$\sigma$} (see main text, section IV.B). Thus our explicit result
for the propagator of the reduced density matrix is now
\begin{align}
{\cal K}_{jj'}^{ll'}(t) \;=\;  \sum_{pp'} \langle \langle e^{i (\mu
+ N\bar{p}) \mbox{\boldmath $\alpha$}\cdot\mbox{\boldmath $\sigma$}}
\rangle \rangle^{N_s}  \; e^{-i\bar{p}\Phi}e^{-i\Phi \mu/N}i^{N\bar{p} +
\mu} J^{}_{Np+j-l}(2\Delta^{}_{o}t)J^{}_{Np'+j'-l'}(2\Delta^{}_{o}t)
 \label{K-av2}
\end{align}

To make this result more specific, let us assume the Gaussian
distribution of couplings given in the text (eqtn. (\ref{F_0-bar})).
If we again assume a thermal average then we can easily evaluate
this average, in the same way as in the main text, to get
\begin{align}
\langle \langle e^{i (\mu + N\bar{p}) \mbox{\boldmath
$\alpha$}\cdot\mbox{\boldmath $\sigma$}} \rangle \rangle^{N_s} \;\;
\rightarrow \;\;    \exp[-\lambda(N\bar{p} + \mu)^2/2]
 \label{F-av2}
\end{align}
As before, the result (\ref{K-av2}) can be rewritten as a single
sum, and this is again fairly easily summed numerically. We note
that the effect of the influence function is now to rapidly suppress
paths in which $| N\bar{p} + \mu |$ is non-zero.

\end{widetext}

Now consider the current $I_{j,j+1}(t)$. Again we may distinguish
between the case where we average only over the bath states, and the
case where we average over both the bath states and the bath
couplings. In the case of a single average over bath states, with
fixed couplings, the current is given by Eqs. (\ref{54}) and
(\ref{55}), with the brackets in (\ref{54}) replaced by double
brackets when we average over the bath couplings as well. We see
that formally everything depends only on the phase between sites $j$
and $j+1$, via the bath-generated phase-dependent coupling
$\tilde{\Delta}_{j,j+1}$, and on the density matrix element
$\rho_{j,j+1}(t)$ and its conjugate at time $t$. However this
apparent simplicity is deceptive, because the density matrix depends
itself on the form of the initial density matrix $\rho^{(in)}_{ll'}$
at time $t=0$, and on the propagation of this density matrix in the
interim; thus (\ref{54}) contains implicitly the full propagator
$K_{jj'}^{ll'}(p,p')$.

\begin{widetext}

Using the results derived above for this propagator, we can now
derive expressions for $I_{j,j+1}(t)$. In what follows we only quote
the results of the case of a bath average with fixed couplings - the
case where one also does an ensemble average over the couplings is
easily deduced from these expressions, following the same manoeuvres
as above. The results can be found in both single and double winding
number forms. The double Bessel function form is
\begin{align}
I_{j,j+1}(t)\;=\; 2&\Delta_o \sum_{pp'}
  J_{Np+j-l}(2\Delta_o t) J_{Np'+j+1-l'}(2\Delta_o t) \nonumber \\
 & \times\mbox{Re}\langle \rho^{(in)}_{ll'}i^{N(p-p')}e^{i[(p-p') +
 {1 \over N}]\Phi}\; e^{-i (p-p')\sum_k
 \sum_{ \langle mn\rangle=\langle 01\rangle}^{\langle N0\rangle}
  \mbox{\boldmath $\alpha$}_k^{ mn}\cdot\mbox{\boldmath $\sigma$}_k}
e^{2i\sum_k \sum_{ \langle mn\rangle=\langle
j',j'+1\rangle}^{\langle j-1,j\rangle}\mbox{\boldmath
$\alpha$}_k^{j, j+1}\cdot\mbox{\boldmath $\sigma$}_k}\rangle
\end{align}
Again, let us make the assumption of a completely ring-symmetric
bath, so that $\mbox{\boldmath $\alpha$}_k^{ij} \rightarrow
\mbox{\boldmath $\alpha$}_k$. Then we get
\begin{align}
 \label{eq:I_nn1}
I_{j,j+1}(t) \;=\; 2\Delta_o\sum_{pp'}\sum_{l,l'}
J_{Np+j-l}(2\Delta_o t) J_{Np'+j+1-l'}(2\Delta_o t)
F_{l,l'}(p',p)\times
\mbox{Re}[\rho^{(in)}_{ll'}e^{i\Phi[p'-p+(l-l')/N)]}]
\end{align}
From this we can derive the single Bessel Function summation form as
follows. Using the equation
\begin{align}
\sum_p  J_{Np+n-l}(x)J_{Np+n-l+\nu}(x)=
\frac{1}{N}\sum_{m=0}^{N-1}J_k(2x\sin\frac{k_m}{2})e^{-i(n-l)k_m-i(k_m-\pi)\nu/2}
 \end{align}
which is another form of Graf's identity\cite{Graf}, we set
$\nu=N(p'-p)+1+l-l'$, $x=2\Delta_o t$; then
\begin{align}
I_{j,j+1}(t)
\;=\;\frac{2\Delta_o}{N}\sum_{m=0}^{N-1}\sum_p\sum_{l,l'}&
J_{Np+1+l-l'}(4\Delta_o t\sin\frac{k_m}{2}) \nonumber\\
& \times \; e^{-ik_m[\frac{Np+1}{2}+n-(l+l')/2]}
i^{Np+1+l-l'}F_{ll'}(p)
\mbox{Re}[\rho^{(in)}_{ll'}e^{i\Phi[(p'-p+l-l')/N)]}]
\end{align}
where we define $F_{ll'}(p,0) \equiv F_{ll'}(p)$.

If we make the assumption that the particle starts at the origin,
these results simplify considerably; one gets
\begin{align}
 \label{eq:I_nn2}
I_{j,j+1}(t) \;=\; &2\Delta_o\sum_{pp'} J_{Np+j}(2\Delta_o t)
J_{Np'+j+1}(2\Delta_o t) F_0(p',p)\cos[(\frac{\pi}{2}N+\Phi)(p'-p)]
 \nonumber\\
=\;&\frac{2\Delta_o}{N}\sum_{m=0}^{N-1}\sum_p J_{Np+1}(4\Delta_o
t\sin\frac{k_m}{2})
e^{-ik_m(\frac{Np+1}{2}+j)}i^{Np+1}F_0(p)\cos[(\frac{\pi}{2}N+\Phi)p]
\end{align}
\end{widetext}
for the double and single sums over winding numbers, respectively;
and $F_0(p) \equiv F_{jj}(p,0)$. The latter expression is used in
the text for practical analysis.


\begin{thebibliography}{9}


\bibitem{thouless}     D.J. Thouless, "{\it Topological Quantum
                         numbers in non-relativisitc physics}",
                         World Scientific (1998)

\bibitem{orbital}      L. Pauling, J. Chem. Phys. {\bf 4}, 673
                         (1936); F. London, J. Phys. Radium {\bf 8}, 397 (1937)


\bibitem{LH-2}         H. Lee, Y.-C. Cheng, G.R. Fleming, Science {\bf 316},
                          1462 (2007); see also A. Damjanovic, I. Kosztin, U. Kleinekathofer, K. Schulten, Phys. Rev.
                          {\bf E65}, 031919 (2002), and X. Hu, K. Schulten,
                          Phys. Today {\bf 50} (8), 28 (1997)

\bibitem{AhB-scond}    Y. Imry, "{\it Introduction to Mesoscopic
                          Physics}", Oxford University Press (1997)

\bibitem{QIP}           The earliest quantum ideas for quantum
                          computation involved 'control loops' (see,
                          eg., R.P. Feynman, Found. Phys. {\bf 16},
                          507 (1986); or D. de Falco, D. Tamascelli,
                          J. Phys {\bf A37}, 909 (2004)). The role of
                          loops in modern quantum information
                          processing is most clearly seen in the
                          quantum walk formulation - see E. Farhi,
                          S. Gutmann, Phys. Rev. {\bf A58}, 915 (1998); J. Kempe,
                          Contemp. Phys. {\bf 44}, 307 (2003);
                          A.P. Hines, P.C.E. Stamp, Phys. Rev.
                          {\bf A75}, 062321 (2007); and also refs.
                          \cite{PS06,QW-diss} below.

\bibitem{tau-phi}      P. Mohanty, E. M. Q. Jariwala, R. A. Webb, Phys.
                         Rev. Lett. {\bf 78}, 3366 (1997); J. von
                         Delft, pp. 115-138 in "{\it Fundamental
                         Problems of Mesoscopic Physics}", ed. I.V.
                         Lerner, B.L. Altshuler, Y. Gefen (Kluwer,
                         2004); and refs. therein.

\bibitem{TLS-TauPhi}    F. Pierre, N.O. Birge, Phys. Rev. Lett. {\bf 89},
                        206804 (2002); F.Pierre {\it et al.}, Phys. Rev. {\bf B68},
                        085413 (2003)

\bibitem{TLS-SQUID}     J.M. Martinis {\it et al.}, Phys.Rev. Lett. {\bf 95},
                           210503 (2005)

\bibitem{SHPMP}         P.C.E. Stamp, Stud. Hist. Phil. Mod. Phys. {\bf
                            37}, 467 (2006)

\bibitem{PS00}          N.V. Prokof'ev, P.C.E. Stamp, Rep. Prog.
                            Phys. {\bf 63}, 669 (2000)

\bibitem{feynman63}     R.P. Feynman, F.L. Vernon, Ann. Phys. (NY)
                           {\bf 24}, 118 (1963)

\bibitem{cal83}         A.O. Caldeira, A.J. Leggett, Ann. Phys.
                           (NY), {\bf 149}, 374 (1983)

\bibitem{weiss99}       U. Weiss, "{\it Quantum Dissipative Systems}", World
                           Scientific (1999)

\bibitem{AJL84}         A.J. Leggett, Phys. Rev. {\bf B30}, 1208 (1984)

\bibitem{cal84}         A.O. Caldeira, A.J. Leggett, Physica
                           {\bf 121A}, 587 (1983)

\bibitem{gaita08}       P.C.E. Stamp, A. Gaita-Ari\~{n}o, J.
                           Mat. Chem. {\bf 19}, 1718 (2009)

\bibitem{guinea}        See F. Guinea, Phys. Rev {\bf B65}, 205317
                           (2002);  D. Cohen, B. Horovitz,
                           Europhys. Lett. {\bf 81}, 30001 (2008);
                           and refs. therein.

\bibitem{unruh95}       W.G. Unruh, Phys. Rev. {\bf A51}, 992 (1995)

\bibitem{ekert96}       G.M. Palma, K-A. Suominen, A. Ekert, Proc.
                            Roy. Soc. {\bf A452}, 567 (1996)

\bibitem{milburn05}     C.M. Dawson, A.P. Hines, R.H. McKenzie, G.J.
                            Milburn, Phys. Rev. {\bf A71}, 052321
                            (2005)

\bibitem{PS06}          N.V. Prokof'ev, P.C.E. Stamp, Phys. Rev.
                          {\bf A74}, 020102(R) (2006)

\bibitem{QW-diss}       V. Kendon, Math. Struct. Comp. Sci. {\bf 17}, 1169 (2006)

\bibitem{QDot-ring}     O. Entin-Wohlman, Y. Imry, A. G. Aronov, and Y.
                          Levinson, Phys. Rev. {\bf B51}, 11584
                          (1995).

\bibitem{coleman}         C. G. Callan, S. Coleman, Phys. Rev. {\bf
                            D16}, 1762 (1977)

\bibitem{dube98}          M. Dub\'{e}, P.C.E. Stamp, J. Low. Temp. Phys.
                             {\bf 110}, 779-840 (1998).

\bibitem{PS93}            N.V. Prokof'ev, P.C.E. Stamp, J. Phys. CM
                          {\bf 5}, L663 (1993)

\bibitem{TPS96}           I.S. Tupitsyn, N.V. Prokof'ev, P.C.E. Stamp,
                             Int. J. Mod. Phys. B11, 2901-2926 (1997).

\bibitem{ajl84}           A.J. Leggett, Phys. Rev. {\bf B30}, 1208 (1984)

\bibitem{amb83}           V. Ambegaokar, U. Eckern, G. Sch$\ddot{o}$n,
                          Phys. Rev. Lett. {\bf 48}, 1745 (1982).

\bibitem{wada78}          Y. Wada, J.R. Schrieffer, Phys. Rev. {\bf B18}, 3897
                          (1978); P.C.E. Stamp, Phys. Rev. Lett.
                          {\bf 66}, 2802 (1991); A.H. Castro-Neto and A.O.Caldeira,
                          Phys. Rev. {\bf E48}, 4037 (1993); and
                          ref. \cite{dube98}.

\bibitem{zhen3}           Z. Zhu, P.C.E. Stamp, to be published

\bibitem{ajl87}           A.J. Leggett, S. Chakravarty, A.T. Dorsey, M.P.A. Fisher,
                            A. Garg, W. Zwerger, Rev. Mod. Phys.
                            {\bf 59}, 1 (1987)

\bibitem{schmid83}        A. Schmid, Phys. Rev. Lett. {\bf 51}, 1506 (1983)

\bibitem{mpaf85}          M.P.A. Fisher, W. Zwerger, Phys. Rev. {\bf B32},
                            6190 (1985)

\bibitem{Graf}            G.N. Watson, "{\it A Treatise on the
                            Theory of Bessel Functions}", section
                            11.3, pp. 359-361 (Merchant books,
                            2008).


\end{thebibliography}
\end{document}